\documentclass[12pt,letter,english,intoc,bibliography=totoc,index=totoc,BCOR10mm,captions=tableheading,nottitlepage,fleqn]{article}

\usepackage{setspace, geometry}
\usepackage{etex}
\usepackage{babel,newcent,textcomp}

\usepackage{scrextend}
\usepackage{booktabs, threeparttable}

\usepackage[autostyle]{csquotes}
\usepackage[style=bwl-FU, citestyle=authoryear, backend=biber, useprefix, maxbibnames=99]{biblatex}
\usepackage{lmodern}

\usepackage[T1]{fontenc}
\usepackage[latin9]{inputenc}
\usepackage{sgame}
\usepackage{fancyhdr}
\usepackage{amsmath,amssymb,bbm}
\usepackage[amsthm]{ntheorem}
\usepackage{dsfont}
\usepackage{tablefootnote}
\usepackage[labelfont=bf]{caption}
\usepackage{enumitem}
\usepackage{geometry}
\usepackage{lscape}
\usepackage[unicode=true,
 bookmarks=true,bookmarksnumbered=true,bookmarksopen=true,bookmarksopenlevel=1,
 breaklinks=false,pdfborder={0 0 0},backref=false,colorlinks=true,linkcolor=black,citecolor=black,anchorcolor=black,urlcolor=blue]
 {hyperref}
\hypersetup{pdftitle={Your title},
 pdfauthor={Your name},
 pdfpagelayout=OneColumn, pdfnewwindow=true, pdfstartview=XYZ, plainpages=false}

\usepackage{tikz}

\usetikzlibrary{shapes.geometric, arrows, positioning}
\tikzset{
    >=stealth',
    box/.style={
           rectangle,
           rounded corners,
           draw=black, very thick,
           text width=6.5em,
           minimum height=1.5em,
           text centered},
    boxsmall/.style={
           rectangle,
           rounded corners,
           draw=black, very thick,
           text width=6em,
           minimum height=1.5em,
           text centered},
     simple/.style={
           text centered},
    arrow/.style={
           ->,
           thick,
           shorten <=2pt,
           shorten >=2pt,}
}
\usepackage{graphicx}
\graphicspath{{"/Users/michaelthaler/Dropbox/projects/alive/motivated updating/experiment/"}}

\usepackage[multiple]{footmisc}
\usepackage{cleveref}
\usepackage[figure]{hypcap}
\usepackage{calc}

\let\mySection\section\renewcommand{\section}{\suppressfloats[t]\mySection}

\theoremstyle{break}

\makeatother

\begin{document}

\lhead[]{}

\rhead[]{}

\lfoot[\thepage]{}

\cfoot{}

\rfoot[]{\thepage}

\newpage

\begin{titlepage}
\title{Gender Differences in Motivated Reasoning\thanks{\scriptsize{I would like to thank guest editor Lionel Page and two anonymous referees for their helpful suggestions. I also appreciate the valuable feedback I received from Alberto Alesina, Christine Exley, David Laibson, Matthew Rabin, and Mattie Toma. I am grateful for funding support from the Harvard Business School Research Fellowship and the Eric M. Mindich Research Fund for the Foundations of Human Behavior. The experiment was conducted when I was at Harvard University, and was deemed exempt by the IRB at Harvard (IRB17-1725).}}}
\author{Michael Thaler\thanks{\scriptsize{Princeton University. Email: \href{michael.thaler@princeton.edu}{michael.thaler@princeton.edu}.}}}
\date{July 2021}
\vspace{-3mm}

\maketitle

\abstract{Men and women systematically differ in their beliefs about their performance relative to others; in particular, men tend to be more overconfident. This paper provides support for one explanation for gender differences in overconfidence, \textit{performance-motivated reasoning}, in which people distort how they process new information in ways that make them believe they outperformed others. Using a large online experiment, I find that male subjects distort information processing in ways that favor their performance, while female subjects do not systematically distort information processing in either direction. These statistically-significant gender differences in performance-motivated reasoning mimic gender differences in overconfidence; beliefs of male subjects are systematically overconfident, while beliefs of female subjects are well-calibrated on average. The experiment also includes political questions, and finds that politically-motivated reasoning is similar for both men and women. These results suggest that, while men and women are both susceptible to motivated reasoning in general, men find it particularly attractive to believe that they outperformed others.

\vspace{5mm}
}


\bigskip

\setcounter{page}{0}
\thispagestyle{empty}
\end{titlepage}
\pagebreak \newpage


\section{Introduction}

When applying for jobs, entering competitions, or picking stocks to trade, people must predict how their own performance will compare to the performance of others. A growing body of literature shows that there are sizable gender gaps in beliefs about one's relative performance. The consistent finding across labor market, tournament, and finance environments is that men are more overconfident than women (\cite{HK13}; \cite{NV07}; \cite{BO01}).

Information can remedy or exacerbate these gender gaps. On the one hand, informative signals can enable people to learn about their performance, which will reduce overconfidence; in environments in which men are more overconfident, grounding them in the truth will reduce gender differences. On the other hand, people may distort how they process information in directions that make them falsely believe they are high performers: \textit{performance-motivated reasoning}. If there are gender differences in motivated reasoning, information may cause men to become even more overconfident and the gender confidence gap can expand. 

To identify performance-motivated reasoning, I use the experiment from \textcite{T-WP}, which asks people to assess the veracity of a news source that tells them their current median beliefs are higher than the truth or lower than the truth. People first report their median belief about their performance relative to other subjects on a set of factual questions about the US, politics, and current events. Then, they are given a binary message from an unknown news source that either says that their true performance is greater than their median belief or less than their median belief. The news source either reports the truth (``True News'') or reports the opposite of the truth (``Fake News''). Since subjects' medians are elicited, subjects have revealed that they believe the correct answer is equally likely to be greater than or less than their median. Therefore, a Bayesian subject would not infer anything about the veracity of the source from a ``greater than'' message or a ``less than'' message. However, ``greater than'' may evoke positive motivated beliefs for subjects who find it attractive to believe they outperformed others, and motivated reasoning would lead these subjects to assess this source as more likely to be True News. 

I run this experiment online with approximately 1000 participants, and this paper's main finding is that men systematically engage in performance-motivated reasoning and that women do not. That is, men think that news is more likely to be from True News if it tells them they should become even more confident than they currently are (``good news''), whereas women trust good news and bad news about performance equally. The gender gap in performance-motivated reasoning is statistically significant at the $p=0.001$ level. 

There are two possible explanations for the finding that men alone motivatedly reason about performance. The first explanation is that men may be more susceptible to the bias of motivated reasoning, and women may reason in a more Bayesian manner. This explanation would imply that, on questions that evoke motivated beliefs about other topics, men would again engage in a greater degree of motivated reasoning than women would. The second explanation is that men and women may hold different motivated beliefs in the particular domain of relative performance, but that both would be susceptible to motivated reasoning in other settings. 

To test these competing hypotheses, I consider gender differences in motivated reasoning about politics. On political questions, I find that both male and female subjects engage in politically-motivated reasoning, and that the differences between men and women are small and statistically insignificant.\footnote{This form of heterogeneity is also mentioned in passing in \textcite{T-WP}.} This result suggests that susceptibility to motivated reasoning in general need not be different between men and women, but that certain types of motivated beliefs differ by gender (such as in \cite{CCK-WP}). It is consistent with men finding it more attractive to believe that they outperformed others, but both men and women finding it attractive to believe that their political party's stances are right. 

The findings of gender differences in motivated reasoning about performance may relate to evolutionary theories of overconfidence and self-deception such as those in \textcite{vHT11}, which posit that people deceive themselves in order to better deceive others. \textcite{B88} finds related evidence that men trying to attract women tend to overstate their accomplishments more (possibly due to traditional tendencies for men to try to impress women more than vice versa), and both \textcite{SvdW19} and \textcite{SKPvH20} find experimental evidence that being expected to persuade others increases one's own overconfidence. Performance about (political) knowledge may therefore be a setting that is particularly ripe to lead to gender differences in motivated reasoning. Using the classification of \textcite{CCK-WP}, these results suggest that performance in this setting is more male-typed. 

The data suggest that results are not driven by potential confounding factors of subjects misreporting their median belief or subjects misunderstanding the distribution of news sources. In particular, subjects who have a confidence interval that is symmetric about their initial guess are likely to have a belief distribution for which their mean and median are similar. For such \textit{unskewed} subjects, I find that the treatment effects are nearly identical to the full sample. For the distribution of news sources, I tell some subjects that the (ex ante) likelihood of receiving True News and Fake News is 50-50; the other subjects are not given this information. The first group may anchor more towards 50-50, and the second group may update their meta-beliefs about the distribution; however, I find no significant differences in treatment effects between the two groups, suggesting that these potential confounds are not a principal driver of the main treatment effects.

This paper contributes to a decades-old literature that discusses why there have been substantial gender differences in labor market outcomes (\cite{G90}; \cite{G14}). As shown by numerous experiments (\cite{GNR03}; \cite{NV07}; \cite{DF11}) and the literature summarized in \textcite{NV11}, men tend to engage in competition more than women do. Both preferences (\cite{EG08}; \cite{CG09}; \cite{SE18}) and beliefs about one's ability and performance play a role. In particular, gender differences in overconfidence may be explained by responses to new information (\cite{L77}; \cite{B90}; \cite{BO01}), leading to differential behavior. This paper focuses on a form of overconfidence often described as \textit{overplacement} (\cite{MH08}), in which people are overconfidence about the ranking of their performance or ability within a comparison group; overplacement is particularly important in competitive settings. These findings also complement the empirical results from \textcite{SX21}, who find survey evidence of gender gaps in perceived knowledge --- relating both to overconfidence and overprecision --- about economic questions.

Motivated reasoning from new information is an important potential driver of overconfidence (\cite{K90}; \cite{BT02}), but has been difficult to experimentally identify from other models of cognition (\cite{B19}; \cite{LHM18}; \cite{TPR20}). As such, I use the experimental design from \textcite{T-WP}, which is able to portably identify motivated reasoning in many domains since there is nothing to infer from the signals subjects receive. Using designs that estimate non-Bayesian responsiveness to ``good news'' and ``bad news,'' previous papers have not systematically found statistically significant gender differences in asymmetric updating, a primary consequence of motivated reasoning (\cite{ER11}; \cite{MNNR-WP}; \cite{E11}; \cite{C18}).\footnote{\textcite{E11} does find gender differences in asymmetric updating on a verbal task, but not on a math task.} However, heterogeneity by gender is not the main focus of those papers, and they are not well-powered to rule out moderate differences. In fact, several papers using such designs have found null effects of motivated reasoning more broadly (\cite{B20}; \cite{BGvdW18}), suggesting that these designs struggle to precisely detect the effect. In addition, as argued by \textcite{T-WPb}, ``good news'' is not motivatedly reasoned about when it does not pertain to ones self-image or identity, an effect that appears for both men and women. It is worth mentioning that \textcite{MNNR-WP} and \textcite{C18} find that women underreact more to information than men do, while \textcite{BGvdW18} do not; this dimension is not captured using my design, which rules out underreaction by giving subjects uninformative signals. 

One related paper with a similar hypothesis is \textcite{CCK-WP}, which ran contemporaneous experiments to the one in this paper, and finds that gender differences in responses to information about performance depend on whether the task is stereotypically male-typed or female-typed. My paper differs in its primary focus on identifying motivated reasoning, which the new experimental design enables me to cleanly identify. While \textcite{CCK-WP} find novel evidence for asymmetries in responses to information, most of the effect is due to Bayesian updating or to inaccurate prior beliefs.\footnote{\textcite{CCK-WP} also test a version of the regression specification of \textcite{ER11}. Once they control for Bayesian predictions, they find a suggestive, but statistically insignificant, effect of gender-congruent motivated reasoning.}

The results from my paper indicate that differences in performance-motivated reasoning may be an important determinant of differences in overconfidence. Indeed, I find that overconfidence and motivated reasoning are correlated. Previous literature has looked at potentially-related implications of these gender differences, such as in labor markets (\cite{S-WP}; \cite{ST16}; \cite{SE18}), self-promotion (\cite{EK-WP}), giving in dictator games (\cite{K18}), and stereotyping others (\cite{BCGS19}; \cite{GEKZ19}). This paper adds to this literature by suggesting a driver of persistent differences in overconfidence.

The rest of the paper proceeds as follows: \Cref{design} presents the theory of motivated reasoning and the experimental design. \Cref{results} analyzes experimental results, including the evidence for gender differences in overconfidence and motivated reasoning, the differences between motivated reasoning about performance and politics, and the robustness tests described above. \Cref{conclusion} concludes and proposes directions for future work.

\section{Theory and Experiment Design}
\label{design}

\subsection{Theory}

This section outlines the model of motivated reasoning (as introduced in \cite{T-WP}). I will use the model to introduce the signal structure in the experiment detailed in following sections, compare Bayesian updating to motivated reasoning, and discuss how gender differences in motivated reasoning can manifest themselves.

The premise of the theory is that agents distort how they process new information by acting as if they receive an additional signal corresponding to their motivated beliefs. When signals are uninformative, as will be the case in the experiment, this distortion leads motivated reasoners to update in the direction that they find more attractive. 

There are a set of agents $i$ and a set of questions $q$. For each agent and question, Nature determines whether a source is true ($T$) or false ($\neg T$). Sources are independently drawn with $P(T) = p$, and agents receive data $x_{iq}$ about the source veracity. 

Agents engage in \textit{motivated reasoning}, which means that they form their posterior by incorporating prior, likelihood, and a motivated beliefs term: 
\begin{equation*}
\underbrace{\mathbb{P}(T|x_{iq})}_{\text{posterior}} \propto \underbrace{\mathbb{P}(T)}_{\text{prior}} \cdot \underbrace{\mathbb{P}(x_{iq}|T)}_{\text{likelihood}} \cdot \underbrace{M_{iq}(T) ^ {\varphi_i}}_{\text{mot. reasoning}},
\end{equation*}
We take log odds ratios of both sides to attain an additive form:
\begin{equation}
\label{moreas-eq}
\text{logit }\mathbb{P}(T|x_{iq}) = \text{logit }\mathbb{P}(T) + \log \left(\frac{\mathbb{P}(x_{iq}|T)}{\mathbb{P}(x_{iq}|\neg T)}\right) + \varphi_i (m_{iq}(T) - m_{iq}(\neg T)).
\end{equation}
Motivated reasoners act as if they receive both the actual signal ($x_{iq}$) and a signal whose relative likelihood corresponds to how much they are motivated to believe the state is $T$. $m_{iq}(T): \{T, \neg T\} \to \mathbb{R}$ is denoted the \textbf{motive} function. The weight put on this signal is $\varphi_i \geq 0$, called \textbf{susceptibility}. An agent with $\varphi_i = 0$ is Bayesian; an agent with $\varphi_i > 0$ motivatedly reasons. 

We particularly consider one information structure, in which the $x_{iq}$ are binary and uninformative about the news source veracity. That is, we will choose the $x_{iq}$ in such a way that $\mathbb{P}(x_{iq}|T) = \mathbb{P}(x_{iq}|\neg T) = 1/2$. In such a setting, described in the experiment below, the motivated reasoner will have $\text{logit }\mathbb{P}(T|x_{iq}) = \text{logit }\mathbb{P}(T) + \varphi_i (m_{iq}(T) - m_{iq}(\neg T)).$ If the message is attractive, then their posterior on the likelihood that the news source is true will be higher; if the message is unattractive, then their posterior will be lower. On the other hand, a Bayesian will not update from either message. 

This paper will allow for two types of agents, male and female, who may differ in their motives and their susceptibility: male agents have motives $m_{\text{Male, } q}(T)$ and susceptibility $\varphi_{\text{Male}} \geq 0$, and female agents have motives $m_{\text{Female, } q}(T)$, and susceptibility $\varphi_{\text{Female}} \geq 0$. For each gender type $g$, the procedure backs out $\varphi_g \cdot (m_{g, q}(T) - m_{g, q}(\neg T))$ by observing agents' inference processes.

As further discussed when results are presented, we will observe that $\varphi_g > 0$ for both men and women using evidence from political questions. We will also see evidence that $\varphi_{\text{Male}} \cdot (m_{\text{Male, } q}(T) - m_{\text{Male, } q}(\neg T)) > 0$, while $\varphi_{\text{Female}} \cdot (m_{\text{Female, } q}(T) - m_{\text{Female, } q}(\neg T)) \approx 0$ when the news reports ``good news'' on a question about performance. This evidence is consistent with $m_{\text{Male, } \text{Perf}}(T) > m_{\text{Male, } \text{Perf}}(\neg T)$ and $m_{\text{Female, } \text{Perf}}(T) \approx m_{\text{Female, } \text{Perf}}(\neg T)$. That is, this set of results can be explained by men and women having similar levels of susceptibility and having similar magnitudes of motivated beliefs about politics, but having different motivated beliefs about performance.

\subsection{Outline of Design}

To test the model, the experiment provides people with signals about the veracity of news sources that tell them their median beliefs are higher than the truth or lower than the truth (\cite{T-WP}). These signals are constructed such that there would be nothing for a Bayesian to infer, but they evoke motivated beliefs. 

First, subjects take a study that tests knowledge and responses to news about political and US knowledge issues like crime, income mobility, racial discrimination, and geography. This part of the study is designed to measure both knowledge and politically-motivated reasoning. Full question texts are in the Appendix. Next, subjects are asked the following question: 

\vspace{8mm}
\begin{addmargin}[1cm]{1cm}
\textit{How well do you think you performed on this study about political and U.S. knowledge? I've compared the average points you scored for all questions (prior to this one) to that of 100 other participants.}

\vspace{4mm}
\noindent \textit{How many of the 100 do you think you scored higher than?}

\vspace{4mm}

\noindent \textit{(Please guess between 0 and 100.)}

\end{addmargin}
\vspace{8mm}

\noindent Subjects were indeed ranked versus 100 others, so this question has a factual answer. Subjects may be motivated to believe that they outperformed others, and we will hypothesize that there are gender differences in performance-motivated reasoning. This test of motivated reasoning involves three steps:

\begin{enumerate}
\item \textbf{Beliefs:} Subjects are asked to guess the answers to the question above. They are asked and incentivized to guess their median belief (such that they find it equally likely for the answer to be above or below their guess). Details on incentives are below.

\item \textbf{News:} Subjects receive a binary message from one of two randomly-chosen news sources: True News and Fake News. The message from True News is always correct, and the message from Fake News is always incorrect. This is the main treatment variation.

\hspace{5mm} The message says either ``The answer is \textbf{greater than} your previous guess of [previous guess].'' or ``The answer is \textbf{less than} your previous guess of [previous guess].'' Note that the exact messages are different for subjects who make different initial guesses. 

\hspace{5mm} For the question above, ``greater than'' corresponds to Pro-Performance News, or Good News; ``less than'' corresponds to Anti-Performance News, or Bad News. 

\item \textbf{Assessment:} After receiving the message, subjects assess the probability that the message came from True News using a scale from 0/10 to 10/10, and are incentivized to state their true belief. I test for motivated reasoning by looking at the treatment effect of seeing a ``greater than'' message versus a ``less than'' message on news veracity assessments. 
\end{enumerate}
An equivalent procedure is followed for political questions participants answer elsewhere in the study. Each subject sees nine political questions and one performance question; on each of these questions, their median belief is elicited and they are asked to assess a news source that says the answer is greater than or less than this belief.\footnote{Eight political questions are seen before the performance question; one question is seen after it.} On the political questions, the news is either classified as Pro-Republican or Pro-Democratic depending on the topic; in these cases, we use subjects' stated party preference to classify the news as either being Good News / Pro-Party News or Bad News / Anti-Party News.

Recall that since subjects receive ``greater than'' or ``less than'' messages that compare the answer to their median, a Bayesian would not change her assessment based on the message.\footnote{Likewise, general over- and under-weighting of priors and likelihoods (such as forms of prior-confirming biases and conservatism) do not predict a treatment effect of message direction on assessment.} If she had a prior that P(True News) = 1/2 before seeing the message, she would form a posterior that P(True $|$ ``greater than'') = P(True $|$ ``less than'') = 1/2. We attribute systematic treatment effects of the messages on veracity assessments to motivated reasoning. For instance, if men tend to state P(True $|$ ``greater than'') $>$ P(True $|$ ``less than'') and women tend to state P(True $|$ ``greater than'') $=$ P(True $|$ ``less than'') on the question about performance, this would indicate that there is systematic performance-motivated reasoning for men but not for women.

\subsection{Pages and Scoring Rules}
\label{experiment-pages}

\noindent \textbf{Overall Scoring Rule}

At the end of the experiment, subjects earn a show-up fee of \$3 and either receive an additional \$10 bonus or no additional bonus. As will be elaborated below, in each round of the experiment subjects earn between 0-100 ``points'' based on their performance. These points correspond to the probability that the subject wins the bonus: a score of $x$ points corresponds to an $x/10$ percent chance of winning the bonus.\footnote{This lottery system is a form of the binarized scoring rule designed to account for risk aversion, as directly mapping points to earnings could lead to subjects strategically hedging their guesses (\cite{HO13}). As such, we do not need to assume risk neutrality in order for the experiment to be incentive compatible, but we do need to assume that subjects behave as if compound lotteries are reduced to simple lotteries. The probability distribution is identical to randomly choosing a question for payment and subsequently playing the lottery based on the points scored on that question.}

\bigskip
\noindent \textbf{Questions Page}

On question pages, subjects are given the text of the question and are asked to input three numbers about their initial beliefs:
\begin{itemize}
\item \textit{My Guess}: This elicits the median of the subjects' prior distribution.
\item \textit{My Lower Bound}: This elicits the 25th percentile of the subjects' prior distribution.
\item \textit{My Upper Bound}: This elicits the 75th percentile of the subjects' prior distribution.
\end{itemize}

The scoring rule for guesses is piecewise linear. Subjects earn $\max \{100-|c-g|, 0\}$ points for a guess of $g$ when the correct answer is $c$. Subjects are told that they will maximize expected points by stating the median of their belief distribution.

The scoring rule for bounds is piecewise linear with different slopes. For upper bound $ub$, subjects earn $\max \{100-3(c-ub), 0\}$ points if $c \geq ub$ and $\max \{100-(ub-c), 0\}$ points if $c \leq ub$. For lower bound $lb$, subjects earn $\max \{100-(c-lb), 0\}$ points if $c \geq lb$ and $\max \{100-3(lb-c), 0\}$ points if $c \leq lb$. Subjects maximize expected points by setting $ub$ to be the $75^\text{th}$ percentile and $lb$ to be the $25^\text{th}$ percentile of their belief distribution. Subjects are restricted to give answers for which My Lower Bound $\leq$ My Guess $\leq$ My Upper Bound; if they do not, they see an error message. 

\bigskip
\noindent \textbf{News Assessments Page}

After submitting their guess, subjects see a second page about the same question. At the top of the page is the exact text of the original question. Below the original question is a message relating the answer to the number they submit for \textit{My Guess}. This message says either: 

\bigskip
``The answer is \textbf{greater than} your previous guess of \textbf{[My Guess]}.'' or

\bigskip
``The answer is \textbf{less than} your previous guess of \textbf{[My Guess]}.''

\bigskip
Subjects are told that True News \textit{always} tells the truth and Fake News \textit{never} tells the truth, and that sources are independent across questions. Below the message, subjects are asked: ``Do you think this information is from True News or Fake News?'' and choose one of eleven radio buttons that say ``x/10 chance it's True News, (10-x)/10 chance it's Fake News'' from each x=0, 1, \dots, 10.

The scoring rule for assessments is quadratic. For assessment $a$, subjects earn $100(1 - (1-a)^2)$ points if the source is True News and $100(1 - a^2)$ points if it is Fake News. Subjects maximize expected points by stating the closest multiple of 0.1 to their belief. Subjects are given a table with the points earned as a function of each assessment and news type.

Occasionally, a subject will correctly guess the answer. If this happens, they skip the news assessment page and moves on to the next question.\footnote{This is true except for the comprehension check question, where the message says ``The answer is \textbf{equal} / \textbf{not equal} to your previous guess of \textbf{[My Guess]}.''} Seven answers to the performance question (0.71 percent) are exactly correct. The likelihood of getting the answer exactly correct is similar to the likelihood if subjects guessed randomly (0.99 percent), so there is no evidence of guess manipulation.

\subsection{Experiment Details}
\label{details}

The experiment, also the focus of \textcite{T-WP}, was conducted in June 2018 on Amazon's Mechanical Turk (MTurk) platform. MTurk is an online labor marketplace in which participants choose ``Human Intelligence Tasks'' to complete. MTurk has become a very popular way to run economic experiments, and participants generally tend to have more diverse demographics than students in university laboratories (e.g. \cite{HRZ11}; \cite{LFD16}). The experiment was coded using oTree, an open-source software based on the Django web application framework developed by \textcite{CSW16}. 

The study was offered to MTurk workers currently living in the United States. 1,387 subjects were recruited and answered at least one question, and 1,300 subjects completed the study. Of these subjects, 987 (76 percent) passed simple attention and comprehension checks, and the rest are dropped from the analyses.\footnote{In order to pass these checks, subjects needed to perfectly answer the comprehension check question in \Cref{comprehension-question} (by giving a correct answer, correct bounds, and answering the news assessment with certainty). In addition, many questions had clear maximum and minimum possible answers (such as percentages, between 0 and 100). Subjects were dropped if any of their answers did not lie within these bounds.} As discussed in \Cref{robustness} and the Appendix, results are robust to the inclusion of these subjects.

Of the 987 subjects, 980 (99.3 percent) do not get the performance question exactly correct. These subjects receive a message relating their guess to the true answer, and are given the news veracity assessment page. Of the 980, 528 (53.9 percent) identify as male, 447 (45.6 percent) identify as female, and 5 (0.5 percent) do not identify as either or prefer not to report their gender. These results focus on the 975 subjects who report their gender as male or female. 

The balance table for the Pro-Performance / Anti-Performance treatment for these 975 subjects is in \Cref{balance-table}. There are not statistically significant differences across demographics between subjects who receive Good News or Bad News about their performance. The experiment is also balanced across news type overall, with 488 (50.1 percent) Good News messages and 487 (49.9 percent) Bad News messages. No subjects exited the experiment after seeing the message for this question, so selective attrition is not a concern.

When looking at behavior on politicized questions, we consider subjects' subjective ratings of the Republican Party and Democratic Party. In the total sample, 627 subjects (64 percent) give a higher rating to the Democratic Party; 270 (27 percent) give a higher rating to the Republican Party; and 90 (9 percent) give identical ratings to both parties. These subjects are labeled as ``Pro-Dem,'' ``Pro-Rep,'' and ``Neutral,'' respectively, and for analyses that include political questions the Neutral subjects are dropped.

Political differences between men and women in the sample are modest; women are slightly more likely to prefer the Democratic Party. 62 percent of men are Pro-Dem and 28 percent are Pro-Rep, while 66 percent of women are Pro-Dem and 27 percent are Pro-Rep. Male and female party preferences are not statistically significantly different from each other in this sample.

\section{Results}
\label{results}

\Cref{overconfidence-section} analyzes gender differences in overconfidence about performance. Next, \Cref{moreas-section} presents the main results about how men and women motivatedly reason about their performance. It shows that gender gaps in motivated reasoning on the performance question is not mirrored on politicized questions. \Cref{robustness} discusses robustness and potential confounds to identification, and does not find evidence that results are explained by these alternative explanations. \Cref{discussion} proposes an interpretation of these results, as well as complementary evidence regarding overprecision.

\subsection{Overconfidence}
\label{overconfidence-section}
Gender differences in overconfidence are apparent from the raw data. Male subjects expect to outperform 55.3 percent of subjects (s.e. 0.9 pp), and they actually outperform 49.5 percent of subjects (s.e. 1.1 pp). Female subjects expect to outperform 44.5 percent of subjects (s.e. 1.0 pp), and they actually outperform 45.4 percent of subjects (s.e. 1.3 pp). Male subjects' performance predictions are 5.8 pp too high on average (s.e. 1.3 pp, $p<0.001$), while female subjects' performance predictions are 0.8 pp too low on average (s.e. 1.5 pp, $p = 0.585$). In other words, male subjects are systematically overconfident and female subjects are not. The difference between the overconfidence of male and female subjects is 6.6 pp (s.e. 2.0 pp, $p<0.001$).

\begin{figure}[htb!]
\caption{Confidence and Performance by Gender}
\begin{center}
\vspace{-5mm}
\includegraphics[width = .9\textwidth]{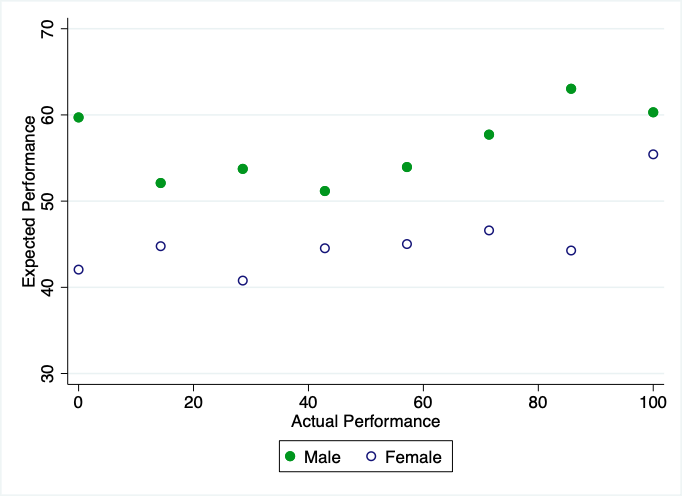}
\end{center}
\label{overconfidence-gender}
\begin{threeparttable}
\begin{tablenotes}
\begin{scriptsize}
\vspace{-10mm}
\item \textbf{Notes:} The graph shows a binned scatter plot of performance and confidence by gender. The x-axis measures performance percentile on the quiz. The y-axis measures confidence, which is what percentile subjects expected to score. 
\end{scriptsize}
\end{tablenotes}
\end{threeparttable}
\end{figure}

Subjects' beliefs about their performance are positively correlated with their true performance, but there is little heterogeneity in the gender gap in overconfidence by true performance. The gender gap is seen among high-, medium-, and low-performing subjects. \Cref{overconfidence-gender} uses a binned scatter plot to compare men and women at various points in the performance distribution. 

The figure shows that confidence is larger for men at every point in the performance distribution on this question. In fact, men of high, medium, and low performance levels all expect to perform at or above the median on average. Except at the high end, women do not expect to perform above the median. The differences between the beliefs of men and women (10.8 pp) are larger than the differences between the beliefs of the top quintile and bottom quintile of subjects (6.3 pp).

Appendix \Cref{overconfidence-gender-cdf} presents a CDF of subjects' levels of overconfidence. It shows that differences in overconfidence are not driven by outliers; the distribution for men first-order-stochastically dominates the distribution for women. 

\subsection{Motivated Reasoning}
\label{moreas-section}

The gender gap in motivated reasoning is also apparent in the raw data. 

\begin{figure}[htb!]
\caption{Trust in News About Own Performance by Gender}
\begin{center}
\vspace{-5mm}
\includegraphics[width = .9\textwidth]{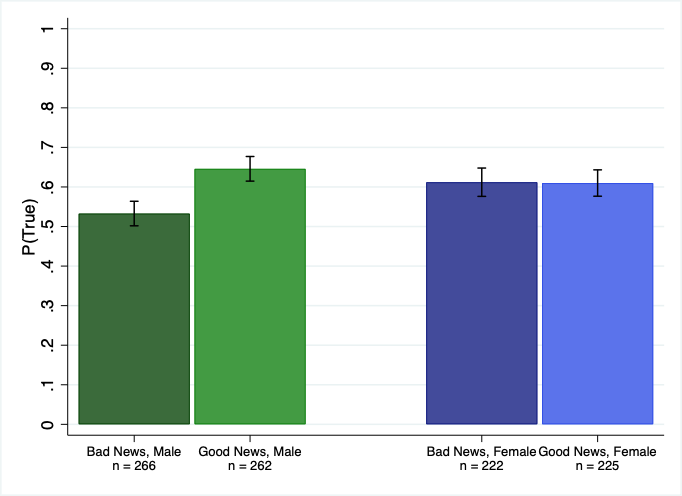}
\end{center}
\label{raw-data-gender}
\begin{threeparttable}
\begin{tablenotes}
\begin{scriptsize}
\vspace{-10mm}
\item \textbf{Notes:} Good News tells subjects that the correct answer is greater than their median beliefs about their performance; Bad News tells subjects the correct answer is less than these beliefs. The y-axis measures subjects' assessments of the veracity of the source. Bayesians would have the same trust in news for Good News and Bad News, and the residual is motivated reasoning. Error bars correspond to 95 percent confidence intervals. 
\end{scriptsize}
\end{tablenotes}
\end{threeparttable}
\end{figure}

Recall that a Bayesian would believe that the news source were equally likely to be True or Fake if it gave good news or bad news about their relative performance. As shown in \Cref{raw-data-gender}, men significantly trust the news more when it gives Good News about their relative performance than when it gives Bad News. By contrast, women give nearly-identical assessments of Good News and Bad News about their performance.

These results demonstrate that, when given news about their relative performance, men on average motivatedly reason to think they did even better than their (already-overconfident) current beliefs, while women are approximately Bayesian on average.\footnote{Note that this is the measure of the \textit{average treatment effect.} The average treatment for women could be due to women not engaging in any motivated reasoning about performance, or due to women being equally likely to motivatedly reason to believe they performed better than expected and believe they performed worse than expected.}

Gender differences in motivated reasoning appear in both directions: men believe that Good News is more likely to be True News and believe that Bad News is more likely to be Fake News. The differences in average treatment effects are not driven by male and female subjects differently choosing extreme probabilities. Appendix \Cref{good-news-gender-cdf} shows the CDF of news veracity assessments P(True News $|$ Good News) and \Cref{bad-news-gender-cdf} shows the CDF of news veracity assessments P(True News $|$ Bad News) about performance. Male subjects give higher veracity assessments of Good News about performance --- and lower veracity assessments of Bad News about performance --- at all points in the distribution.

Next, we ask whether this discrepancy is due to the specific domain of performance or due to an overall susceptibility to motivated reasoning. To tease apart these competing hypotheses, we consider gender differences in motivated reasoning in another setting: politics. The political topics used follow \textcite{T-WP}, and the list of hypothesized political motives is in Appendix \Cref{political-motives}. \Cref{moreas-gender} shows that that there is no sizable heterogeneity by gender in motivated reasoning about politics, suggesting that the gender differences we see are particular to beliefs about performance.

\begin{figure}[htb!]
\caption{Motivated Reasoning by Gender}
\begin{center}
\vspace{-5mm}
\includegraphics[width = .95\textwidth]{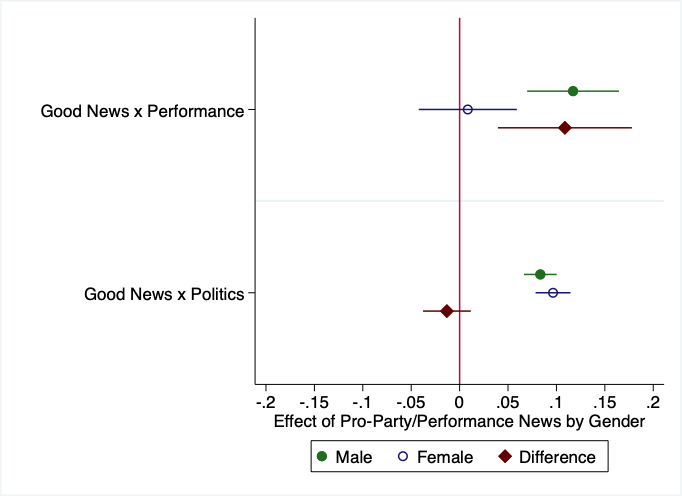}
\end{center}
\label{moreas-gender}
\begin{threeparttable}
\begin{tablenotes}
\begin{scriptsize}
\vspace{-10mm}
\item \textbf{Notes:} The x-axis is the treatment effect of seeing Good News versus Bad News about one's performance or party. Bayesians would have a treatment effect of zero, while motivated reasoners would have a positive effect. Error bars correspond to 95 percent confidence intervals. 
\end{scriptsize}
\end{tablenotes}
\end{threeparttable}
\end{figure}

The regression specifications for gender differences in performance-motivated reasoning are between subjects, regressing assessments $a$ for subject $i$ on the subject's gender, whether the news was Good (vs. Bad) about Performance, and the interaction between gender and news direction, with and without controls for a vector of other demographics $Z_i$.
\begin{equation}
\label{spec1}
a_i = \alpha + \beta_1 \cdot \text{Male}_i + \beta_2 \cdot \text{Good x Performance}_i + \beta_3 \cdot \text{Good x Performance}_i \cdot \text{Male}_i + \gamma Z_i + \epsilon_i
\end{equation}
$\beta_3$ is the main object of interest, measuring how much men motivatedly reason more than women about performance. $\beta_2$ is also interesting in its own right; it measures how much women motivatedly reason about performance. The demographic controls $Z_i$ are race, age, income, education, religion, and party preference.

The regression specification that compares performance-motivated reasoning to politically-motivated reasoning is within subjects, regressing assessments $a$ by subject $i$ on question $q$ in round $r$ on a dummy for the news being Pro-Performance vs. Anti-Performance, a dummy for the news being overall Good vs. overall Bad (this corresponds to Good News about performance or party vs. Bad News about performance or party), and the interactions between gender and news direction for performance and overall. Since this test is within subjects, we can replace controls with fixed effects for subjects $i$, question topic $q$ interacted with gender, and round $r$ interacted with gender. Note that the performance question is always in the same round, so question and round fixed effects only pertain to the political questions.\footnote{The performance question is always in the same round since the correct answer depends on subjects' answers to the previous questions.} For this analysis, we remove the politically-neutral subjects, since there is no hypothesis as to which way they will motivatedly reason about politics.
\begin{eqnarray}
a_{iqr} &=& \alpha + \beta_1 \cdot \text{Good News}_{iqr} + \beta_2 \cdot \text{Good News}_{iqr} \cdot \text{Male}_i + \beta_3 \cdot \text{Good x Performance}_{iqr} + \nonumber \\ 
&& \beta_4 \cdot \text{Good x Performance}_{iqr} \cdot \text{Male}_i + \gamma FE_i + \delta FE_{\text{gender, }q} + \zeta FE_{\text{gender, }r} + \epsilon_{iqr} \label{spec2}
\end{eqnarray}

\Cref{moreas-gender-table} shows that the regression results show the same patterns as what was captured in the raw data. 
\begin{center}
\begin{threeparttable}[htbp!]
\begin{footnotesize}
\caption{Motivated Reasoning by Gender and Topic}
{
\def\sym#1{\ifmmode^{#1}\else\(^{#1}\)\fi}
\begin{tabular}{l*{4}{c}}
\hline\hline
                    &\multicolumn{1}{c}{(1)}         &\multicolumn{1}{c}{(2)}         &\multicolumn{1}{c}{(3)}         &\multicolumn{1}{c}{(4)}         \\
\hline
Good x Performance  &    0.005         &    0.006         &                  &   -0.088\sym{***}\\
                    &  (0.026)         &  (0.026)         &                  &  (0.028)         \\
Good x Performance x Male \hspace{5mm} &    0.113\sym{***}&    0.114\sym{***}&                  &    0.122\sym{***}\\
                &  (0.035)         &  (0.035)         &                  &  (0.037)         \\
Good x Party        &                  &                  &    0.095\sym{***}&                  \\
                    &                  &                  &  (0.009)         &                  \\
Good x Party x Male &                  &                  &   -0.014         &                  \\
                    &                  &                  &  (0.013)         &                  \\
Good News           &                  &                  &                  &    0.097\sym{***}\\
                    &                  &                  &                  &  (0.009)         \\
Good News x Male    &                  &                  &                  &   -0.013         \\
                    &                  &                  &                  &  (0.013)         \\
Gender x Question FE&       No         &       No         &      Yes         &      Yes         \\
Gender x Round FE   &       No         &       No         &      Yes         &      Yes         \\
Subject FE          &       No         &       No         &      Yes         &      Yes         \\
Gender Control      &      Yes         &      Yes         &       No         &       No         \\
Other Controls      &       No         &      Yes         &       No         &       No         \\
\hline
Observations        &      887         &      887         &     7868         &     8755         \\
\(R^{2}\)           &     0.03         &     0.04         &     0.25         &     0.23         \\
Mean                &    0.601         &    0.601         &    0.574         &    0.577         \\
\hline\hline
\multicolumn{5}{l}{\footnotesize Standard errors in parentheses}\\
\multicolumn{5}{l}{\footnotesize \sym{*} \(p<0.10\), \sym{**} \(p<0.05\), \sym{***} \(p<0.01\)}\\
\end{tabular}
}

\label{moreas-gender-table}
\end{footnotesize}
\begin{tablenotes}
\begin{scriptsize}
\item \textbf{Notes:} OLS, errors clustered at subject level. Dependent variable: subjects' elicited probability that their message came from True News. Only subjects who list male or female as their gender, and subjects who have a party preference, are included. Good News includes both Good x Party and Good x Performance, so specification (4) is comparing Good x Performance to Good x Party. Subject controls: race, age, log(income), years of education, religion, and party preference. 
\end{scriptsize}
\vspace{5mm}
\end{tablenotes}
\end{threeparttable}
\end{center}
In particular, columns (1) and (2) use \Cref{spec1} to show that gender interacts with performance-motivated reasoning, and that only men are systematically biased towards trusting Pro-Performance news. Column (3) shows that men and women are similar at motivatedly reasoning towards trusting Pro-Party news. Column (4) uses Equation \hyperref[spec2]{(3)} to show that the difference between motivated reasoning about politics vs. performance is significantly larger for men than it is for women. Note that Good News is defined as \textit{either} good news about performance or politics news, so the Good x Performance row in column (4) measures the difference between Pro-Performance and Pro-Party news.

\begin{figure}[htb!]
\caption{Motivated Reasoning by Gender and Topic}
\begin{center}
\vspace{-5mm}
\includegraphics[width = .9\textwidth]{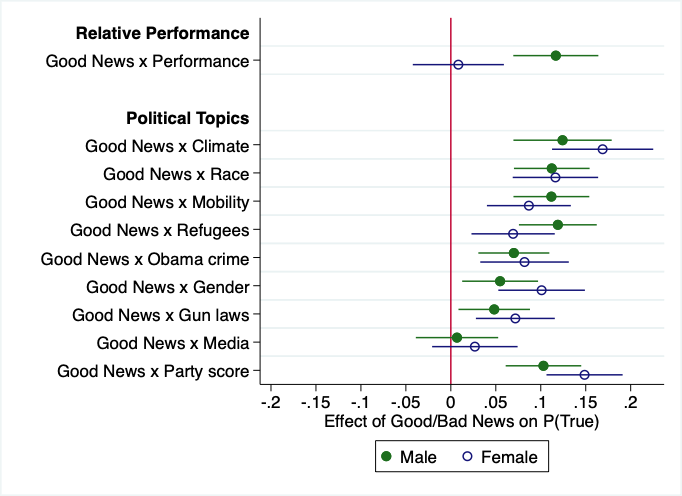}
\end{center}
\label{moreas-gender-topic}
\begin{threeparttable}
\begin{tablenotes}
\begin{scriptsize}
\vspace{-10mm}
\item \textbf{Notes:} The x-axis is the treatment effect of seeing Good News versus Bad News about one's performance or party. Bayesians would have a treatment effect of zero, while motivated reasoners would have a positive effect. Error bars correspond to 95 percent confidence intervals.
\end{scriptsize}
\end{tablenotes}
\end{threeparttable}
\end{figure}

Recall that the experiment tested motivated reasoning on nine separate political questions. Instead of aggregating all the political topics, \Cref{moreas-gender-topic} compares motivated reasoning by gender on each question individually by interacting the motivated reasoning measure with topic-by-topic dummies. 

This figure shows that the null effect in gender differences on politically-motivated reasoning is not driven by large and heterogeneous gender differences on individual questions. On none of the individual political questions do men and women motivatedly reason by a statistically significantly different amount. For both men and women, we can rule out Bayesian updating in favor of politically-driven motivated reasoning on eight of the nine political questions. The performance question uniquely stands out in its gender discrepancy.

\subsection{Robustness}
\label{robustness}

Two threats to identification include misunderstanding the distribution of news sources and misreporting median beliefs in initial guesses.

First, subjects may not understand the distribution of news sources. The actual likelihood of True News and Fake News is 50 percent each. However, subjects who are given this prior may overly anchor towards 50-50, and subjects who are not given a prior may update about the distribution. While it is not clear why there would be gender differences that interact with these effects, I run a between-subjects treatment to ensure that results do not depend on the prior. That is, subjects are either told in the instructions that the news sources were (ex ante) equally likely, or they were not given this information. 

\begin{figure}[htb!]
\caption{Motivated Reasoning by Gender: Robustness} 
\begin{center}
\vspace{-5mm}
\includegraphics[width = .95\textwidth]{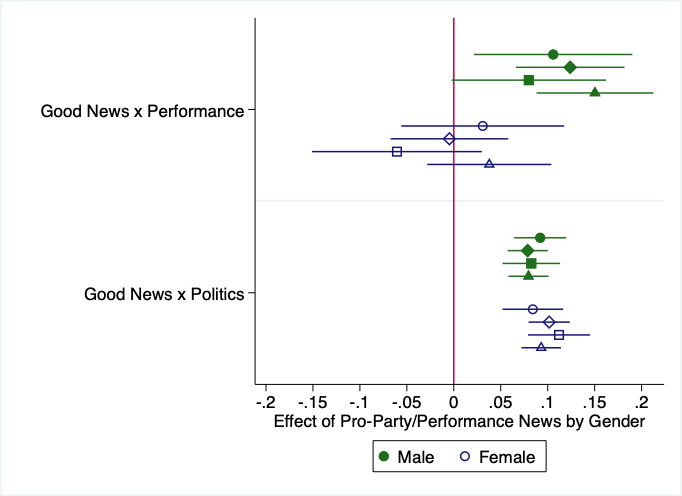}
\end{center}
\label{moreas-gender-robustness}
\begin{threeparttable}
\begin{tablenotes}
\begin{scriptsize}
\vspace{-10mm}
\item \textbf{Notes:} The x-axis is the treatment effect of seeing Good News versus Bad News about one's performance or party. Bayesians would have a treatment effect of zero, while motivated reasoners would have a positive effect. Error bars correspond to 95 percent confidence intervals. Circle: Received 50-50 prior about the veracity of the news. Diamond: Did not receive a prior about the veracity of the news. Square: Unskewed prior belief distributions. Triangle: Skewed prior belief distributions.
\end{scriptsize}
\end{tablenotes}
\end{threeparttable}
\end{figure}

I find that giving subjects a prior about the likelihood of True News does not noticeably affect the results. In each treatment, male subjects systematically engage in performance-motivated reasoning while female subjects do not; and in each treatment, male and female subjects engage in politically-motivated reasoning. These results are shown in the circle and diamond plots in \Cref{moreas-gender-robustness}. In general, there are not clear differences between the treatments, though the effects are noisier. 

Second, subjects may not correctly understand what a median is and report another moment of their belief distribution (such as their mean). If subjects make this error, then results could be explained by Bayesian updating if subjects' mean were lower than their median. Subjects whose initial guesses were lower than their true median belief would rationally think that the source is more likely to be True News if they receive a ``greater than'' message.

To account for this potential confound, I elicit 50-percent confidence intervals (25th and 75th percentile beliefs) from all subjects and analyze where the initial guess lies in this range. If subjects' initial guesses were closer to their lower bound than their upper bound, this would be indicative of negatively-skewed prior belief distributions. I do not find evidence of substantially-skewed distributions. For men, initial guesses lie 49.4 percent of the way between their lower and upper bound; for women, initial guesses lie 51.7 percent of the way between their bounds. Guesses are close to the exact midpoint of confidence intervals, and the gender difference is not large enough to explain the 11 percentage point gap between men's and women's news assessments.

Furthermore, the main results look similar if we restrict estimates to the subjects whose initial guesses lie \textit{exactly} at the midpoint of their confidence interval. As shown in the square and triangle plots in \Cref{moreas-gender-robustness}, there are still sizable gender differences in motivated reasoning on the performance question but not on the political questions.

Results are also similar if subjects who failed attention checks are included in the analysis as well. With the inclusion of the 311 subjects who failed attention checks, there are 1300 subjects in total.\footnote{Most subjects who failed the attention check incorrectly answered the question that asked ``What is the year right now?''} Of these 1300, there are 1282 news veracity assessments on the performance question by men and women; 710 are by men, and 572 are by women. In the Appendix, \Cref{moreas-gender-failed} shows that the results are similar to the results from the main sample (\Cref{moreas-gender}), though the treatment effect estimates are slightly smaller since the inattentive subjects tend to give more random news assessments.

\subsection{Discussion}
\label{discussion}

One explanation for these findings is that, while men and women are both susceptible to biasing how they process information towards more attractive beliefs, they differ in which beliefs they find more attractive. Results suggest that men find it particularly attractive to believe that they outperformed others, while both men and women find it attractive to believe that their politics are right. In the context of the model, these results are consistent with susceptibility to motivated reasoning ($\varphi_g$ in the model) being similar for men and women, but motives ($m_{\text{Male, } q} (\cdot)$ and $m_{\text{Female, } q} (\cdot)$) differing by gender.\footnote{However, note that $\varphi_g$ and $m_{g,q}(\cdot)$ are not separately identified, so both may vary by gender.}

These results provide further evidence that heterogeneity in overconfidence may be partly explained --- and furthered --- by heterogeneity in motivated reasoning. \Cref{moreas-overconfidence-table} provides suggestive evidence that subjects who are more overconfident about their performance also motivatedly reason more about performance. While this evidence is purely correlational, it suggests that motivated reasoning may play a role for overconfidence. 
\begin{center}
\begin{threeparttable}[htbp!]
\begin{footnotesize}
\caption{Motivated Reasoning and Overconfidence about Performance}
{
\def\sym#1{\ifmmode^{#1}\else\(^{#1}\)\fi}
\begin{tabular}{l*{2}{c}}
\hline\hline
                    &\multicolumn{1}{c}{(1)}         &\multicolumn{1}{c}{(2)}         \\
\hline
Good x Performance  &    0.055\sym{***}&    0.056\sym{***}\\
                    &  (0.016)         &  (0.016)         \\
Overconfidence&   -0.015         &   -0.009         \\
                    &  (0.038)         &  (0.037)         \\
Good x Performance x Overconfidence\hspace{5mm} &    0.114\sym{**} &    0.119\sym{**} \\
      &  (0.057)         &  (0.056)         \\
Controls            &       No         &      Yes         \\
\hline
Observations        &      980         &      980         \\
\(R^{2}\)           &     0.02         &     0.04         \\
Mean                &    0.600         &    0.600         \\
\hline\hline
\multicolumn{3}{l}{\footnotesize Standard errors in parentheses}\\
\multicolumn{3}{l}{\footnotesize \sym{*} \(p<0.10\), \sym{**} \(p<0.05\), \sym{***} \(p<0.01\)}\\
\end{tabular}
}

\label{moreas-overconfidence-table}
\end{footnotesize}
\begin{tablenotes}
\begin{scriptsize}
\item \textbf{Notes:} OLS, robust standard errors. Dependent variable: subjects' elicited probability that their message came from True News. Subject controls: gender, race, age, log(income), years of education, religion, and party preference. 
\end{scriptsize}
\vspace{5mm}
\end{tablenotes}
\end{threeparttable}
\end{center}

The main results regarding the issues that men and women differentially motivatedly reason about relate to the broader discussion of gender stereotyping. Results suggest that relative performance is a male-typed belief, while political topics are not systematically gendered. To support the latter hypothesis, we can also look at another measure of confidence: overprecision (\cite{MH08}; \cite{MTH15}; \cite{SX21}). 

In the experiment, subjects are asked to state their 50-percent confidence intervals. On each question, subjects are labeled as \textit{underprecise} if the confidence interval contains the correct answer more than 50 percent of the time, and labeled as \textit{overprecise} if the confidence interval contains the correct answer less than 50 percent of the time.

If gender confidence differences played a role for political beliefs, we would expect men to have greater levels of overprecision than women. However, men and women have almost identical levels of overprecision. Male subjects' confidence intervals include the true answer 46.7 percent of the time (s.e. 0.7 pp) and female subjects' confidence intervals include the true answer 46.6 percent of the time (s.e. 0.8 pp). There is significant evidence of overprecision for both men and women (using a t-test we can reject that these probabilities are less than 50; $p<0.001$ for each men and women).\footnote{\textcite{T-WP} provides a deeper discussion on the relationship between politically-motivated beliefs and overprecision.} The gender difference is 0.1 pp (s.e. 1.1 pp) and statistically insignificant ($p=0.945$). 

The similarities between men and women in political belief overprecision, combined with the previous results on motivated reasoning, suggests that it is the ``outperforming others'' aspect that is the main driver of differences in motivated beliefs about performance. These findings complement the results from \textcite{SX21}, who find survey evidence of gender gaps in perceived knowledge about economic questions.

One potential mechanism for gender differences in motivated reasoning about performance involves the evolutionary basis for overconfidence and self-deception. As theorized by \textcite{vHT11}, people may deceive themselves in order to better persuade others of their ability, and deception may play a particular evolutionary role in mating. For instance, \textcite{B88} finds that men who are trying to attract women tend to brag about their accomplishments more, suggesting that the direction of these gender differences may be more prevalent in societies in which men have traditionally acted to impress women more than women have acted to impress men. Relatedly, men may have also faced more competition in such societies, either due to preferences or traditional gender roles, and have had more incentives to deceive (e.g. \cite{NV07}; \cite{GC07}; \cite{BGGS19}). 

More direct experimental evidence in support of the interpersonal deception hypothesis comes from \textcite{SvdW19} and \textcite{SKPvH20}, who both find that the expectation of having to persuade others leads people to become more overconfident. \textcite{SKPvH20} find related evidence that subjects who will need to persuade others bias how they acquire information, and both papers find that self-deception is effective for interpersonal persuasion. 

From an experimental methodological perspective, these results also suggest that political domains lead to more consistent motivated beliefs than performance domains in the United States today. Researchers who study motivated reasoning may want to consider emphasizing political motives over performance motives if a broader swath of the population is affected by them.

\section{Conclusion}
\label{conclusion}

This paper has shown that there are sizable gender gaps in motivated reasoning. Men systematically engage in motivated reasoning about their performance relative to others; women do not systematically engage in motivated reasoning about performance. The gender gaps in motivated reasoning can make gender gaps in overconfidence persist, and even further them. By contrast, there are little gender differences about politically-motivated reasoning; both men and women are systematically biased in their inference. 

Results are consistent with a theory in which men and women are both susceptible to motivated reasoning, but that there are gender differences in how attractive people find it to believe they performed better than others. 

There are several avenues for future work in both theoretical and applied directions. First, a better understanding of the determinants of motivated beliefs can help us better understand which behavioral differences between men and women are due to beliefs and which are due to preferences. For instance, the causal relationship between preferences for competition and beliefs about relative ability may point in either direction. Men may prefer competitive environments because of their overconfidence, or they may be overconfident about their relative ability because they are motivated to believe that competition will help them. 

Second, these results suggest that debiasing performance-motivated reasoning would reduce gender gaps in overconfidence. Debiasing would be expected to affect men more than women. If motivated reasoning is a cause of overconfidence, men would have a greater reduction in overconfidence. Downstream outcomes of interest --- such as labor market behavior, entry into competitions, and stock trading --- would see a gender convergence as well. Relating the highly structured experimental data from this paper to field evidence would be especially impactful.

\newpage
\nocite{*}
\printbibliography

@article{HRZ11,
		author={John Horton and David Rand and Richard Zeckhauser},
		title={The online laboratory: conducting experiments in a real labor market},
		year={2011},
		journal={Experimental Economics}
}

@article{LFD16,
		author={Kevin Levay and Jeremy Freese and James Druckman},
		title={The Demographic and Political Composition of Mechanical Turk Samples},
		year={2016},
		journal={SAGE Open}
}

@article{BT02,
		author={Roland Benabou and Jean Tirole},
		title={Self-Confidence and Personal Motivation},
		year={2002},
		journal={Quarterly Journal of Economics}
}

@article{BT11,
%		author={Roland Benabou and Jean Tirole},
%		title={Identity, Morals, and Taboos: Beliefs as Assets},
%		year={2011},
%		journal={Quarterly Journal of Economics}
%}

@article{ER11,
		author={David Eil and Justin Rao},
		title={The good news-bad news effect: asymmetric processing of objective information about yourself},
		year={2011},
		journal={American Economic Journal: Microeconomics}
}

@article{MNNR-WP,
		author={Markus Mobius and Muriel Niederle and Paul Niehaus and Tanya Rosenblat},
		title={Managing self-confidence: Theory and experimental evidence},
		year={2014},
		journal={Working Paper}
}

@article{E11,
		author={Seda Ertac},
		title={Does self-relevance affect information processing? Experimental evidence on the response to performance and non-performance feedback},
		year={2011},
		journal={Journal of Economic Behavior and Organization}
}

@article{K14,
% 		author={Camelia Kuhnen},
% 		title={Asymmetric Learning from Financial Information},
% 		year={2014},
% 		journal={The Journal of Finance}
% }

@article{C18,
		author={Alexander Coutts},
		title={Good news and bad news are still news: Experimental evidence on belief updating},
		year={2018},
		journal={Experimental Economics}
}

@article{K90,
		author={Ziva Kunda},
		title={The case for motivated reasoning},
		year={1990},
		journal={Psychological Bulletin}
}

@article{MH08,
		author={Don Moore and Paul Healy},
		title={The Trouble with Overconfidence},
		year={2008},
		journal={Psychological Review}
}

@article{MTH15,
		author={Don Moore and Elizabeth Tenney and Uriel Haran},
		title={Overprecision in Judgment},
		year={2015},
		journal={The Wiley Blackwell Handbook of Judgment and Decision Making}
}

@article{K16a,
% 		author={Dan Kahan},
% 		title={The Politically Motivated Reasoning Paradigm, Part 1: What Politically Motivated Reasoning Is and How to Measure It},
% 		year={2016},
% 		journal={Emerging Trends in Social and Behavioral Sciences}
% }

@article{S-WP,
		author={Heather Sarsons},
		title={Interpreting Signals in the Labor Market: Evidence from Medical Referrals},
		year={2019},
		journal={Working Paper}
}

@article{EK-WPa,
% 		author={Christine Exley and Judd Kessler},
% 		title={Motivated Errors},
% 		year={2018},
% 		journal={Working Paper}
% }

@article{CCK-WP,
		author={Katherine Coffman and Manuela Collis and Leena Kulkarni},
		title={Stereotypes and Belief Updating},
		year={2019},
		journal={Working Paper}
}

@article{CSW16,
		author={Daniel Chen and Martin Schonger and Chris Wickens},
		title={oTree -- An open-source platform for laboratory, online, and field experiments},
		year={2016},
		journal={Journal of Behavioral and Experimental Finance}
}

@article{B19,
		author={Daniel Benjamin},
		title={Errors in Probabilistic Reasoning and Judgment Biases},
		year={2019},
		journal={Chapter for the Handbook of Behavioral Economics}
}

@article{EG16,
%		author={Nicholas Epley and Thomas Gilovich},
%		title={The Mechanics of Motivated Reasoning},
%		year={2016},
%		journal={Journal of Economic Perspectives}
%}

@article{T-WP,
		author={Michael Thaler},
		title={The Fake News Effect: Experimentally Identifying Motivated Reasoning Using Trust in News},
		year={2021},
		journal={Working Paper}
}

@article{T-WPb,
		author={Michael Thaler},
		title={Do People Engage in Motivated Reasoning to Think the World Is a Good Place for Others?},
		year={2020},
		journal={Working Paper}
}

@article{T20R,
%		author={Michael Thaler},
%		title={The ``Fake News'' Effect: Experimentally Identifying Motivated Reasoning Using Trust in News (Registration)},
%		year={2020},
%		journal={AEA RCT Registry},
%		url={https://doi.org/10.1257/rct.4339}
%}

@article{T19R,
%		author={Michael Thaler},
%		title={Debiasing Motivated Reasoning Through Learning: Evidence from an Online Experiment (Registration)},
%		year={2019},
%		journal={AEA RCT Registry},
%		url={https://doi.org/10.1257/rct.4401}
%}

@article{BCGS19,
		author={Pedro Bordalo and Katherine Coffman and Nicola Gennaioli and Andrei Shleifer},
		title={Beliefs about Gender},
		year={2019},
		journal={American Economic Review}
}

@article{EK-WP,
		author={Christine Exley and Judd Kessler},
		title={The gender gap in self-promotion},
		year={2021},
		journal={Working Paper}
}

@article{BO01,
		author={Brad Barber and Terrance Odean},
		title={Boys will be boys: Gender, overconfidence, and common stock investment},
		year={2001},
		journal={Quarterly Journal of Economics}
}

@article{ST16,
		author={Jonathan Schultz and Christian Thoni},
		title={Overconfidence and Career Choice},
		year={2016},
		journal={PLOS One}
}

@article{GEKZ19,
		author={Philip Grossman and Catherine Eckel and Mana Komai and Wei Zhan},
		title={It pays to be a man: Rewards for leaders in a coordination game},
		year={2019},
		journal={Journal of Economic Behavior and Organization}
}

@article{EG08,
		author={Catherine Eckel and Philip Grossman},
		title={Men, Women and Risk Aversion: Experimental Evidence},
		year={2008},
		journal={Handbook of Experimental Economics Results}
}

@article{SE18,
		author={Olga Shurchkov and Catherine Eckel},
		title={Gender Differences in Behavioral Traits and Labor Market Outcomes},
		year={2018},
		journal={The Oxford Handbook of Women and the Economy}
}

@article{CG09,
		author={Rachel Croson and Uri Gneezy},
		title={Gender differences in preferences},
		year={2009},
		journal={Journal of Economic Literature}
}

@article{NV07,
		author={Muriel Niederle and Lise Vesterlund},
		title={Do women shy away from competition? Do men compete too much?},
		year={2007},
		journal={American Economic Review}
}

@article{GNR03,
		author={Uri Gneezy and Muriel Niederle and Aldo Rustichini},
		title={Performance in competitive environments: Gender differences},
		year={2003},
		journal={Quarterly Journal of Economics}
}

@article{DF11,
		author={Thomas Dohmen and Armin Falk},
		title={Performance pay and multidimensional sorting: Productivity, preferences, and gender},
		year={2011},
		journal={American Economic Review}
}

@article{NV11,
		author={Muriel Niederle and Lise Vesterlund},
		title={Gender and Competition},
		year={2011},
		journal={Annual Review of Economics}
}

@article{L77,
		author={Ellen Lenney},
		title={Women's self-confidence in achievement settings},
		year={1977},
		journal={Psychological Bulletin}
}

@article{B90,
		author={Sylvia Beyer},
		title={Gender Differences in the Accuracy of Self-Evaluations of Performance},
		year={1990},
		journal={Journal of Personality and Social Psychology}
}

@book{G90,
		author={Claudia Goldin},
		title={Understanding the Gender Gap: An Economic History of American Women},
		year={1990},
		publisher={Oxford University Press}
}

@article{G14,
		author={Claudia Goldin},
		title={A Grand Gender Convergence: Its Last Chapter},
		year={2014},
		journal={American Economic Review}
}

@article{LHM18,
		author={Jennifer Logg and Uriel Haran and Don Moore},
		title={Is overconfidence a motivated bias? Experimental evidence},
		year={2018},
		journal={Journal of Experimental Psychology}
}

@article{SX21,
		author={Heather Sarsons and Guo Xu},
		title={Confidence Men? Gender and Confidence: Evidence among Top Economists},
		year={2021},
		journal={AEA Papers and Proceedings}
}

@article{K18,
		author={David Klinowski},
		title={Gender differences in giving in the Dictator Game: the role of reluctant altruism},
		year={2018},
		journal={Journal of the Economic Science Association}
}

@article{B20,
		author={Kai Barron},
		title={Belief updating: does the `good-news, bad-news' asymmetry extend to purely financial domains?},
		year={2020},
		journal={Experimental Economics}
}

@article{BGvdW18,
		author={Thomas Buser and Leonie Gerhards and Jo\"{e}l van der Weele},
		title={Responsiveness to feedback as a personal trait},
		year={2018},
		journal={Journal of Risk and Uncertainty}
}

@article{HK13,
		author={Jiekun Huang and Darren Kisgen},
		title={Gender and corporate finance: Are male executives overconfident relative to female executives?},
		year={2013},
		journal={Journal of Financial Economics}
}

@article{vHT11,
		author={William von Hippel and Robert Trivers},
		title={The evolution and psychology of self-deception},
		year={2011},
		journal={Behavioral and Brain Sciences}
}

@article{B88,
		author={David Buss},
		title={The Evolution of Human Intrasexual Competition: Tactics of Mate Attraction},
		year={1988},
		journal={Journal of Personality and Social Psychology}
}

@article{SvdW19,
		author={Peter Schwardmann and Jo\"{e}l van der Weele},
		title={Deception and self-deception},
		year={2019},
		journal={Nature Human Behavior}
}

@article{SKPvH20,
		author={Alice Solda and Changxia Ke and Lionel Page and William von Hippel},
		title={Strategically delusional},
		year={2020},
		journal={Experimental Economics}
}

@article{HO13,
		author={Tanjim Hossain and Ryo Okui},
		title={The binarized scoring rule},
		year={2013},
		journal={Review of Economic Studies}
}

@article{BGGS19,
		author={Stefanie Brilon and Simona Grassi and Manuel Grieder and Jonathan F. Schulz},
		title={Strategic Competition and Self-Confidence},
		year={2019},
		journal={Working Paper}
}

@article{GC07,
		author={Rosanna Guadagno and Robert Cialdini},
		title={Gender Differences in Impression Management in Organizations: A Qualitative Review},
		year={2007},
		journal={Sex Roles}
}

@article{TPR20,
		author={Ben Tappin and Gordon Pennycook and David Rand},
		title={Thinking clearly about causal inferences of politically motivated reasoning: Why paradigmatic study designs often undermine causal inference},
		year={2020},
		journal={Current Opinion in Behavioral Science}
}

\appendix

\newpage

\section{Additional Figures}
\subsection{Balance Table}
\begin{small}
\begin{center}
\def\sym#1{\ifmmode^{#1}\else\(^{#1}\)\fi}
\begin{tabular}{p{3.5cm}*{4}{c}}
\hline\hline
&\multicolumn{1}{c}{Good News}&\multicolumn{1}{c}{Bad News}&\multicolumn{1}{c}{Good vs. Bad}&\multicolumn{1}{p{2cm}}{\centering p-value}\\
\hline
Male  & 0.545 & 0.538 & 0.007 & 0.824 \\   & (0.023) & (0.023) & (0.032) & \\ 
Female & 0.455 & 0.462 & -0.007 & 0.824 \\   & (0.023) & (0.023) & (0.032) & \\
Age  & 34.830 & 35.780 & -0.950 & 0.172 \\   & (0.482) & (0.501) & (0.695) & \\
White  & 0.779 & 0.735 & 0.044 & 0.113 \\   & (0.019) & (0.020) & (0.027) & \\
Black  & 0.074 & 0.088 & -0.015 & 0.406 \\   & (0.012) & (0.013) & (0.017) & \\
Latino  & 0.051 & 0.078 & -0.027 & 0.089 \\   & (0.010) & (0.012) & (0.016) & \\
Education  & 14.617 & 14.719 & -0.102 & 0.392 \\   & (0.085) & (0.084) & (0.119) & \\
Log(income)  & 10.715 & 10.687 & 0.028 & 0.574 \\   & (0.035) & (0.036) & (0.050) & \\
Religious  & 0.465 & 0.417 & 0.048 & 0.129 \\   & (0.023) & (0.022) & (0.032) & \\
Pro-Republican  & 0.279 & 0.269 & 0.010 & 0.735 \\   & (0.020) & (0.020) & (0.029) & \\
Pro-Democratic  & 0.633 & 0.639 & -0.005 & 0.861 \\   & (0.022) & (0.022) & (0.031) & \\
\hline
\(N\)  & 488 & 487 & 975 & \\
\hline
\hline
\end{tabular}
\end{center}
\end{small}
\label{balance-table}
\begin{footnotesize} \textbf{Notes:} Standard errors in parentheses. Only subjects who list male or female as their gender are included. Good News / Bad News refers to news about relative performance. Education is in years. Religious is 1 if subject is in any religious group. Pro-Republican (Pro-Democratic) is 1 if subject gives a strictly higher rating to the Republican (Democratic) Party. 
\end{footnotesize}

\newpage

\subsection{Overconfidence}
\begin{figure}[htb!]
\caption{CDF of Overconfidence by Gender}
\begin{center}
\vspace{-5mm}
\includegraphics[width = .9\textwidth]{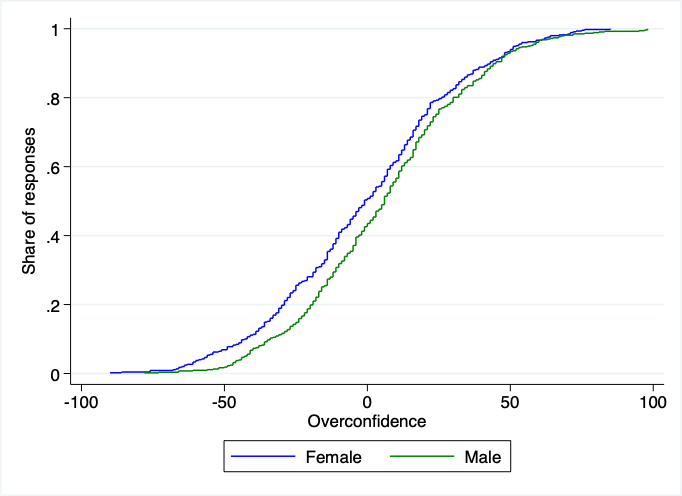}
\end{center}
\label{overconfidence-gender-cdf}
\begin{threeparttable}
\begin{tablenotes}
\begin{scriptsize}
\vspace{-10mm}
\item \textbf{Notes:} Overconfidence is measured by subtracting actual percentile performance (which ranges from 0 to 100) from predicted percentile performance (also ranging from 0 to 100). Positive numbers indicate overconfidence and negative numbers indicate underconfidence.  
\end{scriptsize}
\end{tablenotes}
\end{threeparttable}
\end{figure}

\newpage 

\subsection{Motivated Reasoning}

\begin{figure}[htb!]
\caption{CDF of Trust in ``Good News'' by Gender}
\begin{center}
\vspace{-5mm}
\includegraphics[width = .9\textwidth]{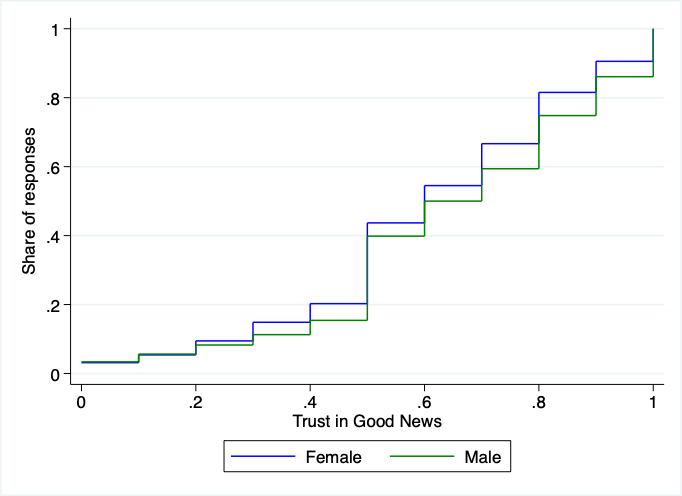}
\end{center}
\label{good-news-gender-cdf}
\begin{threeparttable}
\begin{tablenotes}
\begin{scriptsize}
\vspace{-10mm}
\item \textbf{Notes:} Good News tells subjects that the correct answer is greater than their median beliefs about their performance. This figure shows that men trust Good News more than women do. The x-axis measures subjects' assessments of P(True News $|$ Good News). The y-axis measures the share of respondents that give at most that high of an assessment. Bayesians would have the same trust in news for Good News and Bad News, and the residual is motivated reasoning. Error bars correspond to 95 percent confidence intervals. 
\end{scriptsize}
\end{tablenotes}
\end{threeparttable}
\end{figure}

\newpage 

\begin{figure}[htb!]
\caption{CDF of Trust in ``Bad News'' by Gender}
\begin{center}
\vspace{-5mm}
\includegraphics[width = .9\textwidth]{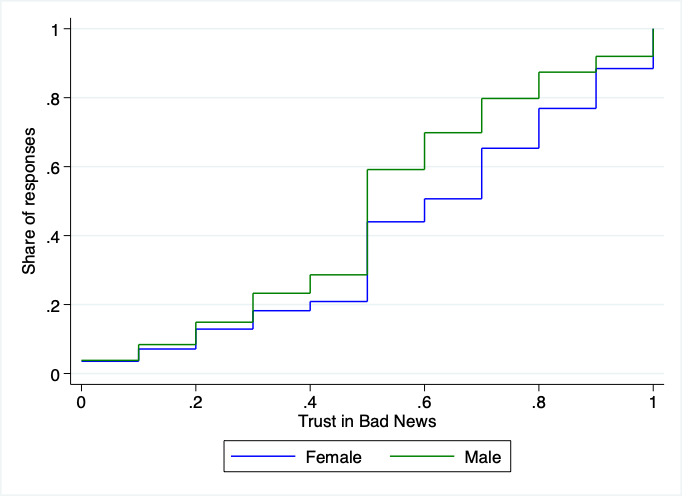}
\end{center}
\label{bad-news-gender-cdf}
\begin{threeparttable}
\begin{tablenotes}
\begin{scriptsize}
\vspace{-10mm}
\item \textbf{Notes:} Bad News tells subjects that the correct answer is less than their median beliefs about their performance. This figure shows that women trust Bad News more than men do. The x-axis measures subjects' assessments of P(True News $|$ Bad News). The y-axis measures the share of respondents that give at most that high of an assessment. Bayesians would have the same trust in news for Good News and Bad News, and the residual is motivated reasoning. Error bars correspond to 95 percent confidence intervals. 
\end{scriptsize}
\end{tablenotes}
\end{threeparttable}
\end{figure}

\newpage 

\renewcommand\arraystretch{1.5}
\begin{table}
\begin{scriptsize}
\caption{Topics and Hypothesized Motives for Democrats and Republicans}
\label{political-motives}
\end{scriptsize}
\begin{tabular}{l l l}  
\toprule
\textbf{Topic} & \textbf{Pro-Democrat Motives} & \textbf{Pro-Republican Motives} \\
\midrule
\hyperref[obama-crime-question]{US crime} & Got better under Obama & Got worse under Obama \\
\hyperref[mobility-question]{Upward mobility} & Low in US after tax cuts & High in US after tax cuts \\
\hyperref[race-question]{Racial discrimination} & Severe in labor market & Not severe in labor market \\
\hyperref[gender-question]{Gender} & Girls better at math & Boys better at math \\
\hyperref[refugees-question]{Refugees} & Decreased violent crime & Increased violent crime \\
\hyperref[global-warming-question]{Climate change} & Scientific consensus & No scientific consensus \\
\hyperref[guns-question]{Gun reform} & Decreased homicides & Didn't decrease homicides \\
\hyperref[media-question]{Media bias} & Media not dominated by Dems & Media is dominated by Dems \\
\hyperref[party-performance-question]{Party performance} & Higher for Dems over Reps & Higher for Reps over Dems \\
\bottomrule
\end{tabular}
\end{table}

\renewcommand\arraystretch{1}

\newpage 

\begin{figure}[htb!]
\caption{Motivated Reasoning by Gender, Including Subjects Who Fail Attention Checks}
\begin{center}
\vspace{-5mm}
\includegraphics[width = \textwidth]{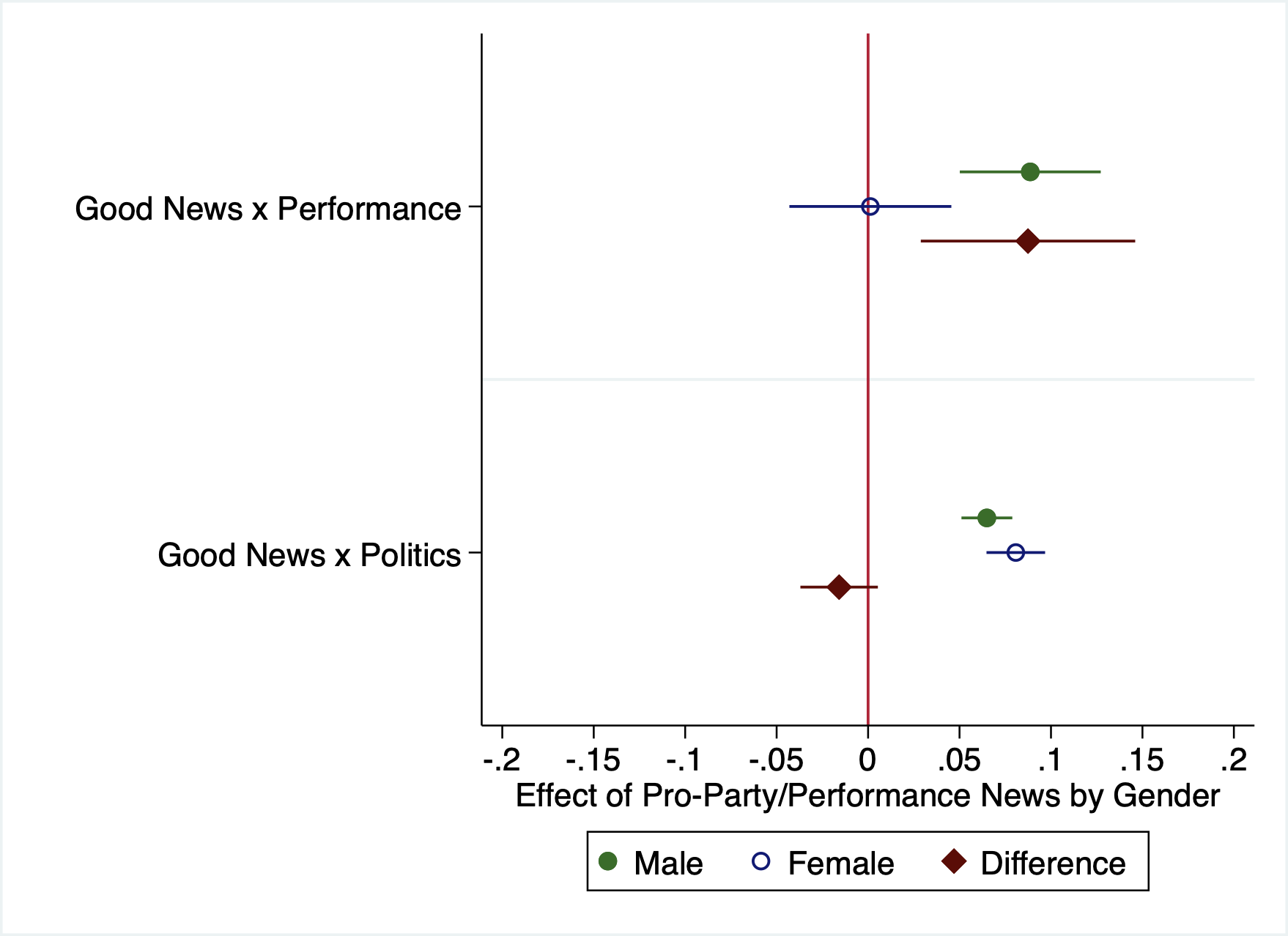}
\end{center}
\label{moreas-gender-failed}
\begin{threeparttable}
\begin{tablenotes}
\begin{scriptsize}
\vspace{-10mm}
\item \textbf{Notes:} Same as \Cref{moreas-gender} but the subjects who failed attention checks are also included. The x-axis is the treatment effect of seeing Good News versus Bad News about one's performance or party. Bayesians would have a treatment effect of zero, while motivated reasoners would have a positive effect. Error bars correspond to 95 percent confidence intervals. 
\end{scriptsize}
\end{tablenotes}
\end{threeparttable}
\end{figure}

\clearpage 

\section{Study Materials: Exact Question Wordings}
\label{question-wordings}

The Relative Performance question is seen in the round before the Party Performance question but after the other political topics.

\begin{small}

\subsection*{Performance Question}
\label{own-performance-question}
How well do you think you performed on this study about political and U.S. knowledge? I've compared the average points you scored for all questions (prior to this one) to that of 100 other participants.

How many of the 100 do you think you scored higher than?

(Please guess between 0 and 100.)

\vspace{2mm}

\textit{Correct answer: Depends on participant's performance.}

\subsection*{Political Questions}
\subsubsection*{Crime Under Obama}
\label{obama-crime-question}
Some people believe that the Obama administration was too soft on crime and that violent crime increased during his presidency, while others believe that President Obama's pushes towards criminal justice reform and reducing incarceration did not increase violent crime.

This question asks how murder and manslaughter rates changed during the Obama administration. In 2008 (before Obama became president), the murder and manslaughter rate was 54 per million Americans.

In 2016 (at the end of Obama's presidency), what was the per-million murder and manslaughter rate?

\vspace{2mm}

\textit{Correct answer: 53.}

\textit{Source linked on results page: \url{http://bit.ly/us-crime-rate}}

\subsubsection*{Upward Mobility}
\label{mobility-question}
In 2017, Donald Trump signed into law the largest tax reform bill since Ronald Reagan's 1981 and 1986 bills. Some people believe that Reagan's reforms accelerated economic growth and allowed lower-income Americans to reap the benefits of lower taxes, while other people believe that this decreased the government's spending to help lower-income Americans get ahead.

This question asks whether children who grew up in low-income families during Reagan's tenure were able to benefit from his tax reforms.

Of Americans who were born in the lowest-income (bottom 20\%) families from 1980-1985, what percent rose out of the lowest-income group as adults?

(Please guess between 0 and 100.)

\vspace{2mm}

\textit{Correct answer: 64.9.}

\textit{Source linked on results page: \url{http://bit.ly/us-upward-mobility} (page 47)}

\subsubsection*{Racial Discrimination}
\label{race-question}
In the United States, white Americans have higher salaries than black Americans on average. Some people attribute these differences in income to differences in education, training, and culture, while others attribute them more to racial discrimination.

In a study, researchers sent fictitious resumes to respond to thousands of help-wanted ads in newspapers. The resumes sent had identical skills and education, but the researchers gave half of the (fake) applicants stereotypically White names such as Emily Walsh and Greg Baker, and gave the other half of the applicants stereotypically Black names such as Lakisha Washington and Jamal Jones.

9.65 percent of the applicants with White-sounding names received a call back. What percent of the applicants with Black-sounding names received a call back?

(Please guess between 0 and 100.)

\vspace{2mm}

\textit{Correct answer: 6.45.}

\textit{Source linked on results page: \url{http://bit.ly/labor-market-discrimination}}

\subsubsection*{Gender and Math GPA}
\label{gender-question}
In the United States, men are more likely to enter into mathematics and math-related fields. Some people attribute this to gender differences in interest in or ability in math, while others attribute it to other factors like gender discrimination.

This question asks whether high school boys and girls differ substantially in how well they do in math classes. A major testing service analyzed data on high school seniors and compared the average GPA for male and female students in various subjects.

Male students averaged a 3.04 GPA (out of 4.00) in math classes. What GPA did female students average in math classes?

(Please guess between 0.00 and 4.00.)

\vspace{2mm}

\textit{Correct answer: 3.15.}

\textit{Source linked on results page: \url{http://bit.ly/gender-hs-gpa}}

\subsubsection*{Refugees and Violent Crime}
\label{refugees-question}
Some people believe that the U.S. has a responsibility to accept refugees into the country, while others believe that an open-doors refugee policy will be taken advantage of by criminals and put Americans at risk.

In 2015, German leader Angela Merkel announced an open-doors policy that allowed all Syrian refugees who had entered Europe to take up residence in Germany. From 2015-17, nearly one million Syrians moved to Germany. This question asks about the effect of Germany's open-doors refugee policy on violent crime rates.

In 2014 (before the influx of refugees), the violent crime rate in Germany was 224.0 per hundred-thousand people.

In 2017 (after the entrance of refugees), what was the violent crime rate in Germany per hundred-thousand people?

\vspace{2mm}

\textit{Correct answer: 228.2.}

\textit{Sources linked on results page: Main site: \url{http://bit.ly/germany-crime-main-site}. 2014 and 2015 data: \url{http://bit.ly/germany-crime-2014-2015}. 2016 and 2017 data: \url{http://bit.ly/germany-crime-2016-2017}.}

\subsubsection*{Climate change}
\label{global-warming-question}
Some people believe that there is a scientific consensus that human activity is causing global warming and that we should have stricter environmental regulations, while others believe that scientists are not in agreement about the existence or cause of global warming and think that stricter environmental regulations will sacrifice jobs without much environmental gain.

This question asks about whether most scientists think that global warming is caused by humans. A major nonpartisan polling company surveyed thousands of scientists about the existence and cause of global warming.

What percent of these scientists believed that ``Climate change is mostly due to human activity''?

(Please guess between 0 and 100.)

\vspace{2mm}

\textit{Correct answer: 87.}

\textit{Source linked on results page: \url{http://bit.ly/scientists-climate-change}}

\subsubsection*{Gun Reform}
\label{guns-question}
The United States has a homicide rate that is much higher than other wealthy countries. Some people attribute this to the prevalence of guns and favor stricter gun laws, while others believe that stricter gun laws will limit Americans' Second Amendment rights without reducing homicides very much.

After a mass shooting in 1996, Australia passed a massive gun control law called the National Firearms Agreement (NFA). The law illegalized, bought back, and destroyed almost one million firearms by 1997, mandated that all non-destroyed firearms be registered, and required a lengthy waiting period for firearm sales.

Democrats and Republicans have each pointed to the NFA as evidence for/against stricter gun laws. This question asks about the effect of the NFA on the homicide rate in Australia.

In the five years before the NFA (1991-1996), there were 319.8 homicides per year in Australia. In the five years after the NFA (1998-2003), how many homicides were there per year in Australia?

\vspace{2mm}

\textit{Correct answer: 318.6.}

\textit{Sources linked on results page: \url{http://bit.ly/australia-homicide-rate} (Suicides declined substantially, however. For details: \url{http://bit.ly/impact-australia-gun-laws}.)}

\subsubsection*{Media Bias}
\label{media-question}
Some people believe that the media is unfairly biased towards Democrats, while some believe it is balanced, and others believe it is biased towards Republicans.

This question asks whether journalists are more likely to be Democrats than Republicans.

A representative sample of journalists were asked about their party affiliation. Of those either affiliated with either the Democratic or Republican Party, what percent of journalists are Republicans?

(Please guess between 0 and 100.)

\vspace{2mm}

\textit{Correct answer: 19.8.}

\textit{Source linked on results page: \url{http://bit.ly/journalist-political-affiliation}}

\subsubsection*{Party Relative Performance}
\label{party-performance-question}
Subjects are randomly assigned to see either the Democrats' score (and asked to predict the Republicans' score) or to see the Republicans' score (and asked to predict the Democrats' score).

\subsubsection*{Democrats' Relative Performance}
This question asks whether you think Democrats or Republicans did better on this study about political and U.S. knowledge. I've compared the average points scored by Democrats and Republicans among 100 participants (not including yourself).

The Republicans scored 70.83 points on average.

How many points do you think the Democrats scored on average?

(Please guess between 0 and 100)

\vspace{2mm}

\textit{Correct answer: 72.44.}

\subsubsection*{Republicans' Relative Performance}
This question asks whether you think Democrats or Republicans did better on this study about political and U.S. knowledge. I've compared the average points scored by Democrats and Republicans among 100 participants (not including yourself).

The Democrats scored 72.44 points on average.

How many points do you think the Republicans scored on average?

(Please guess between 0 and 100)

\vspace{2mm}

\textit{Correct answer: 70.83.}

















\subsection*{Attention Check Question}
\subsubsection*{Current Year}
\label{comprehension-question}
In 1776 our fathers brought forth, upon this continent, a new nation, conceived in Liberty, and dedicated to the proposition that all men are created equal.

What is the year right now?

This is not a trick question and the first sentence is irrelevant; this is a comprehension check to make sure you are paying attention. For this question, your lower and upper bounds should be equal to your guess if you know what year it currently is.

\vspace{2mm}

\textit{Correct answer: 2018.}

\textit{Source linked on results page: \url{http://bit.ly/what-year-is-it}}

\end{small}

\newpage

\newpage

\section{Online Appendix: Study Materials}

\subsection{Flow of Experiment}

Subjects see a series of pages in the following order:
\begin{itemize}
\item Introduction and Consent
\item Demographics and Current Events Quiz
\item Opinions
\item Instructions for Question Pages
\item Question 1
\item Instructions for News Assessment Pages
\item News Assessment 1
\item Question 2, News Assessment 2, \dots, Question 14, News Assessment 14
\item Feedback
\item Results and Payment
\end{itemize}
The Performance question is always in Round 13, and pertains to the performance in Rounds 1-12. As described above, Rounds 1-12 include eight questions on political topics, three questions on ``neutral'' topics (about a random number, and the latitude and longitude of the center of the US), and one attention check.

Screenshots for each of the pages are in the following subsection. Red boxes are not shown to subjects and are included for illustration purposes only. Results pages here are cut off after three questions, but all results are shown to subjects. Choices on the Demographics page are randomly ordered.

\newpage

\subsection{Study Materials}

\begin{center}
\includegraphics[width = \textwidth]{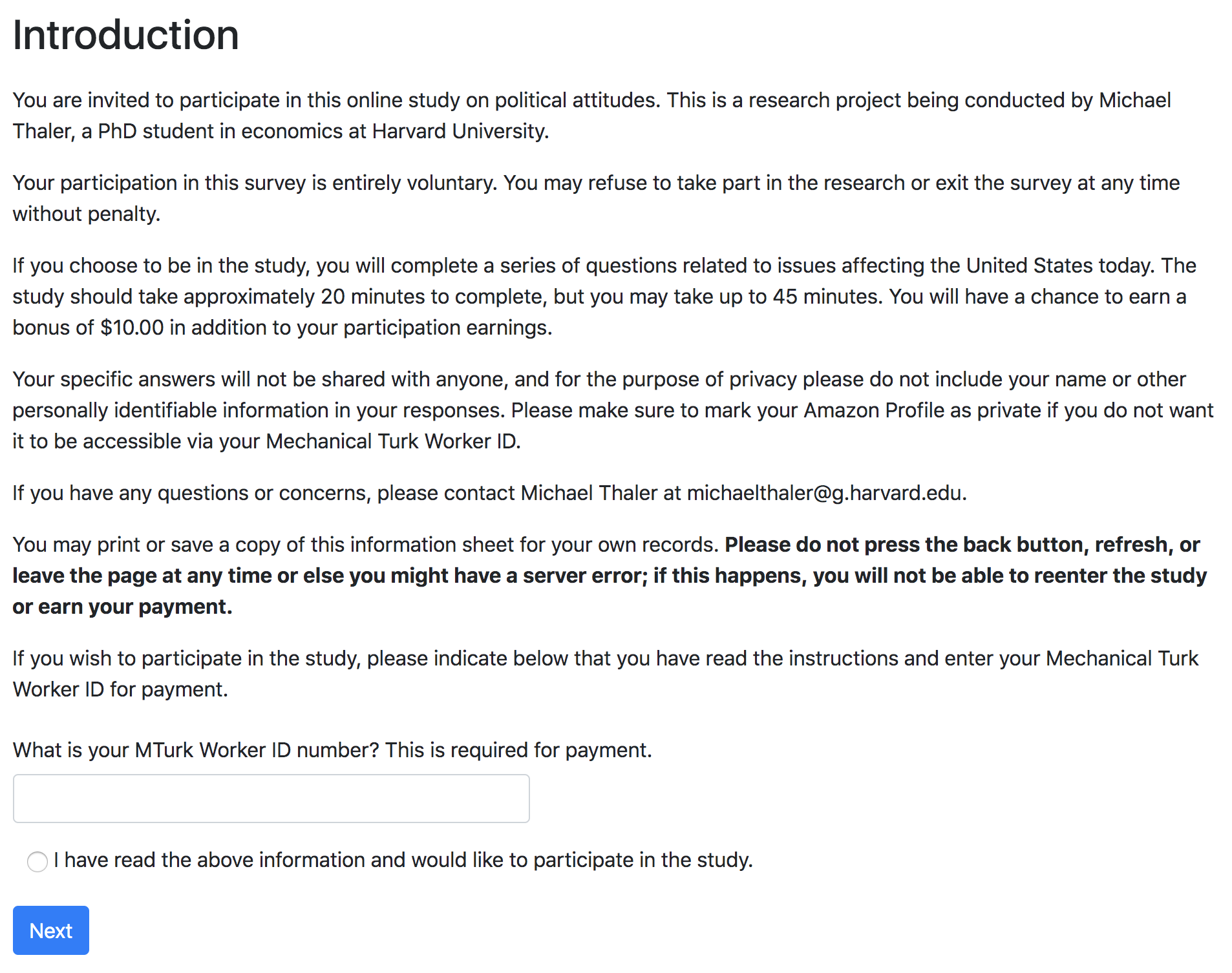}
\end{center}

\newpage

\begin{center}
\includegraphics[height = \textheight]{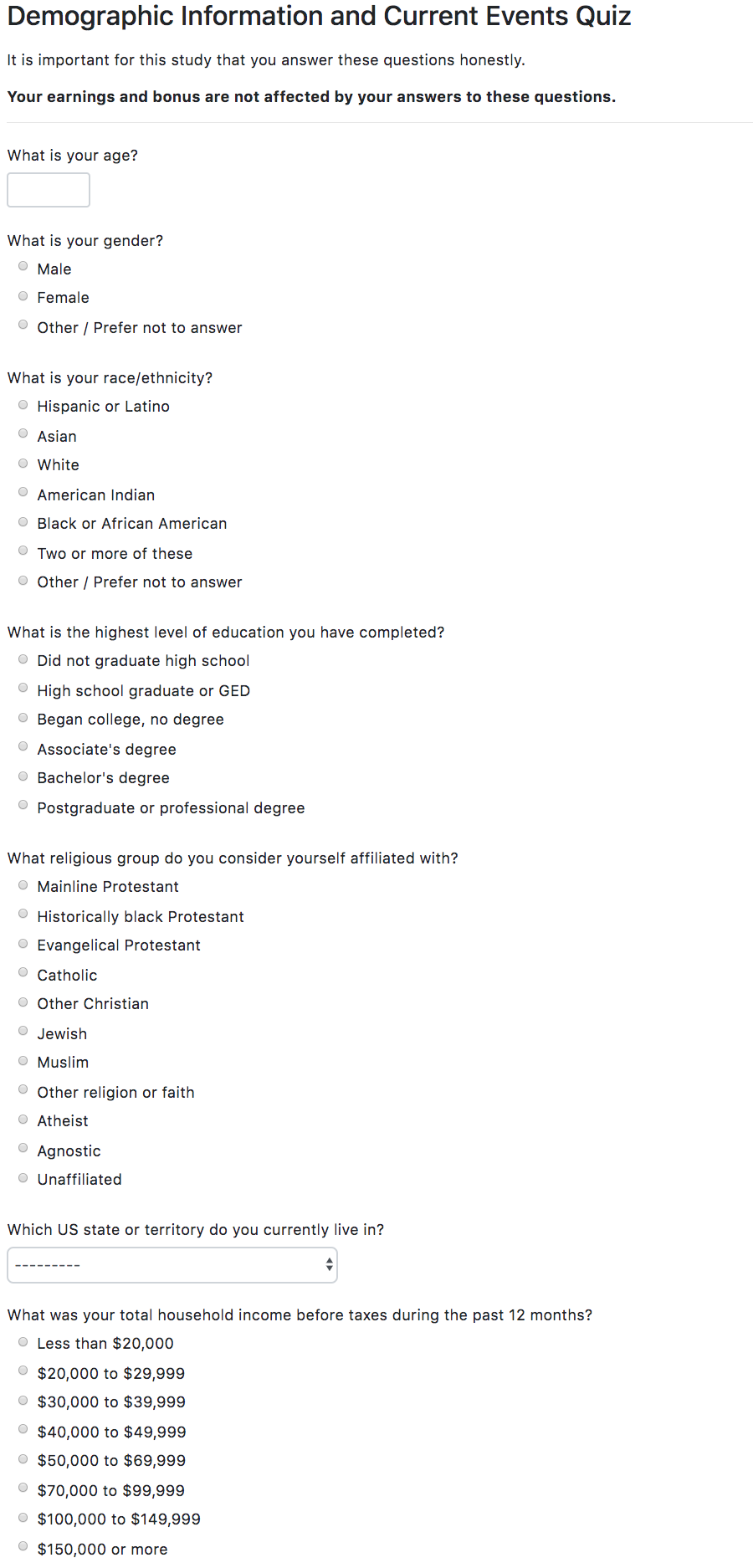}
\end{center}

\newpage

\begin{center}
\includegraphics[height = \textheight]{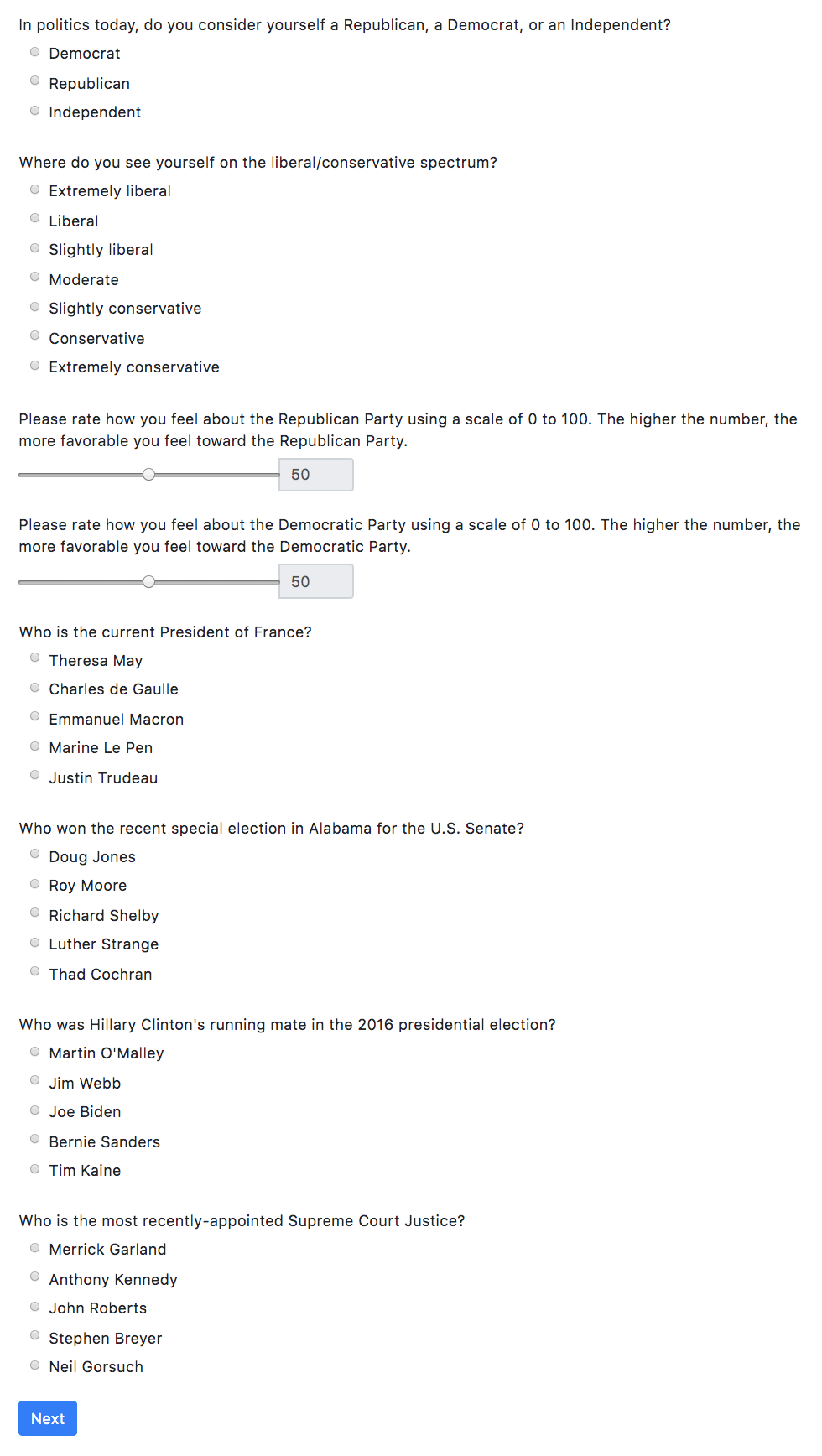}
\end{center}

\newpage


\label{instructions_question_text}
\begin{center}
\includegraphics[width = \textwidth]{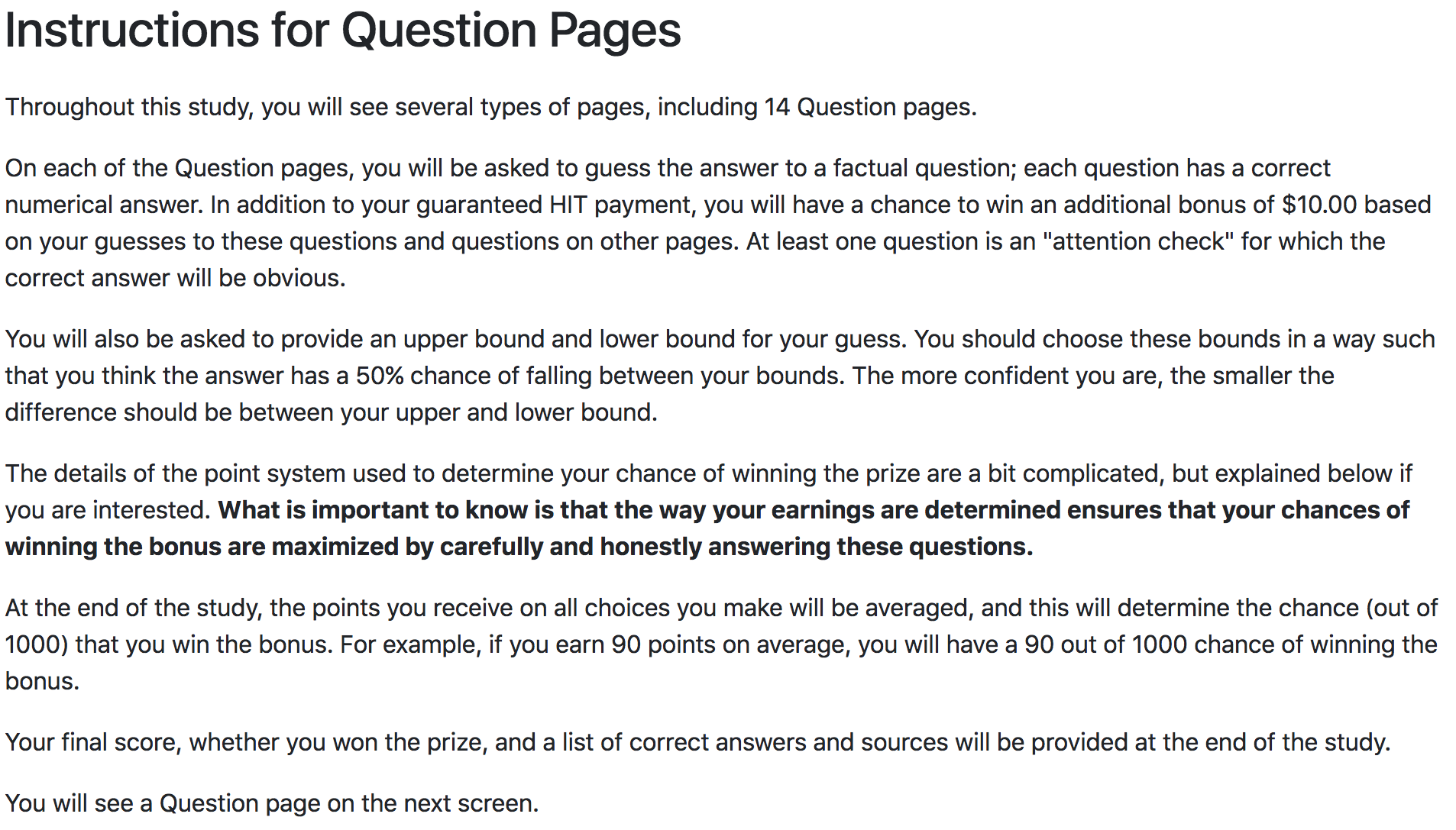}
\end{center}

\label{instructions_question_points}
\begin{center}
\includegraphics[width = \textwidth]{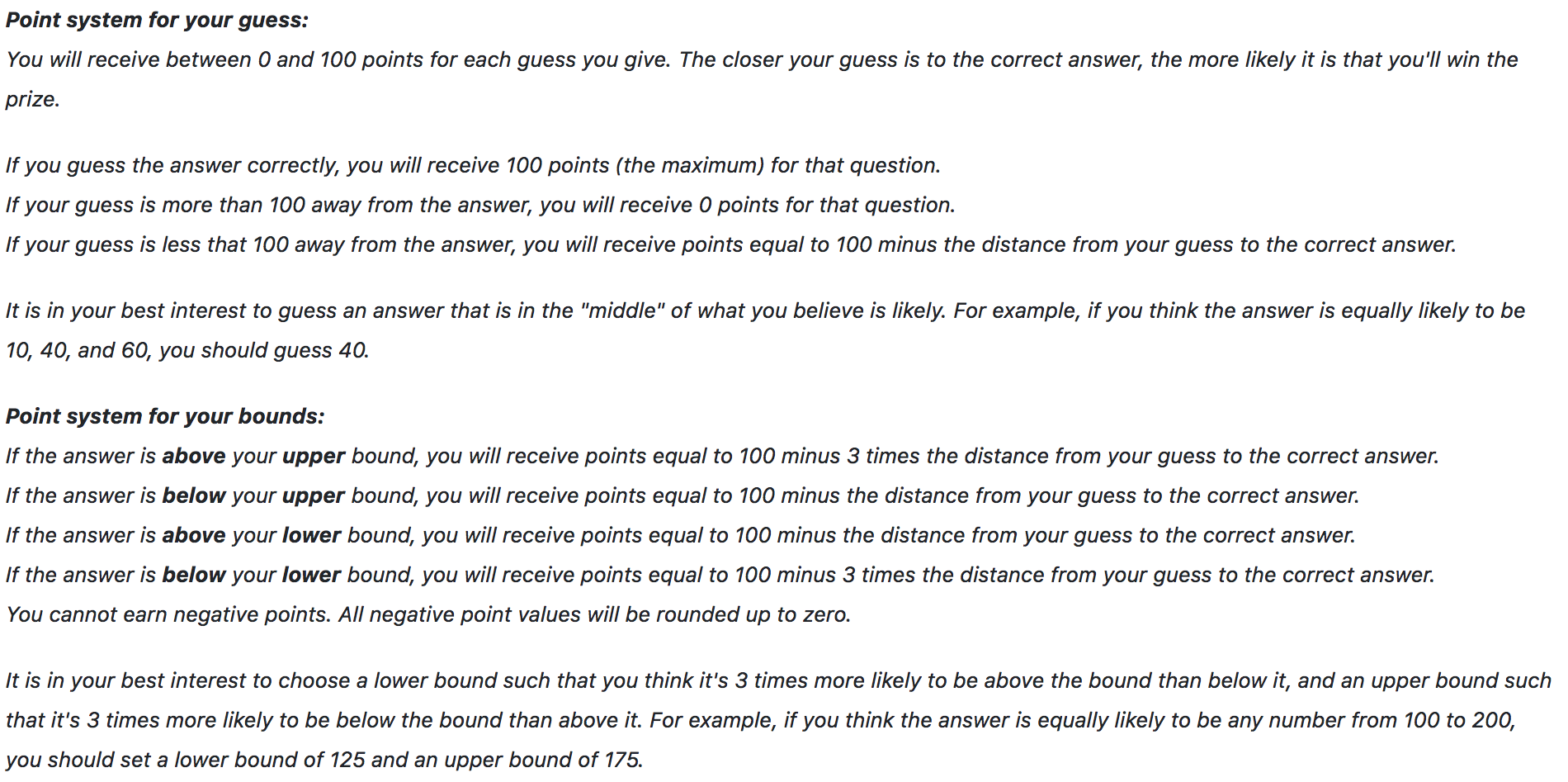}
\end{center}

\newpage


\begin{figure}[h]
\begin{center}
\includegraphics[width = \textwidth]{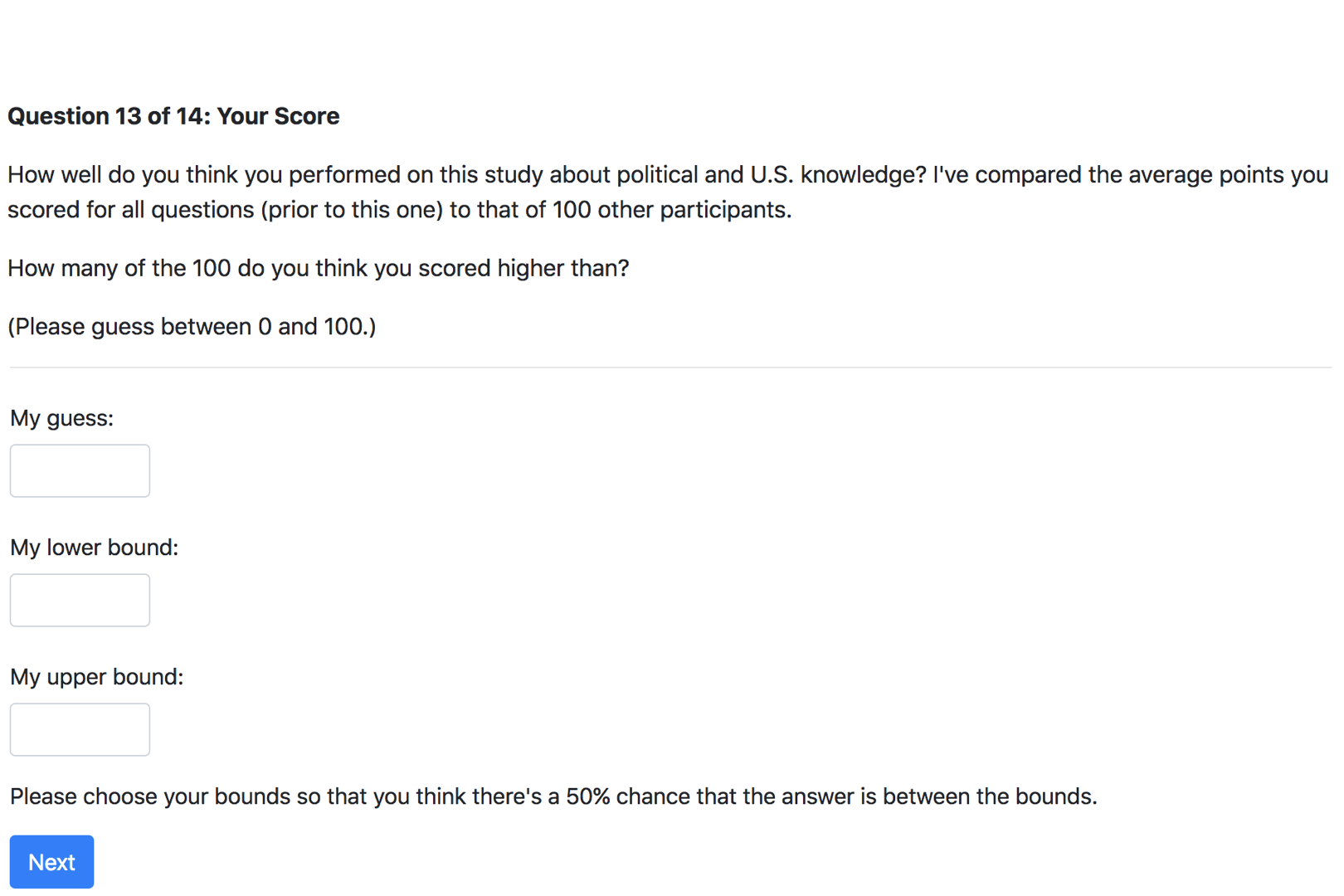}
\end{center}
\caption{The question page for the performance questione.}
\label{question3}
\end{figure}

\newpage


\label{instructions_news_text}
\begin{center}
\includegraphics[width = \textwidth]{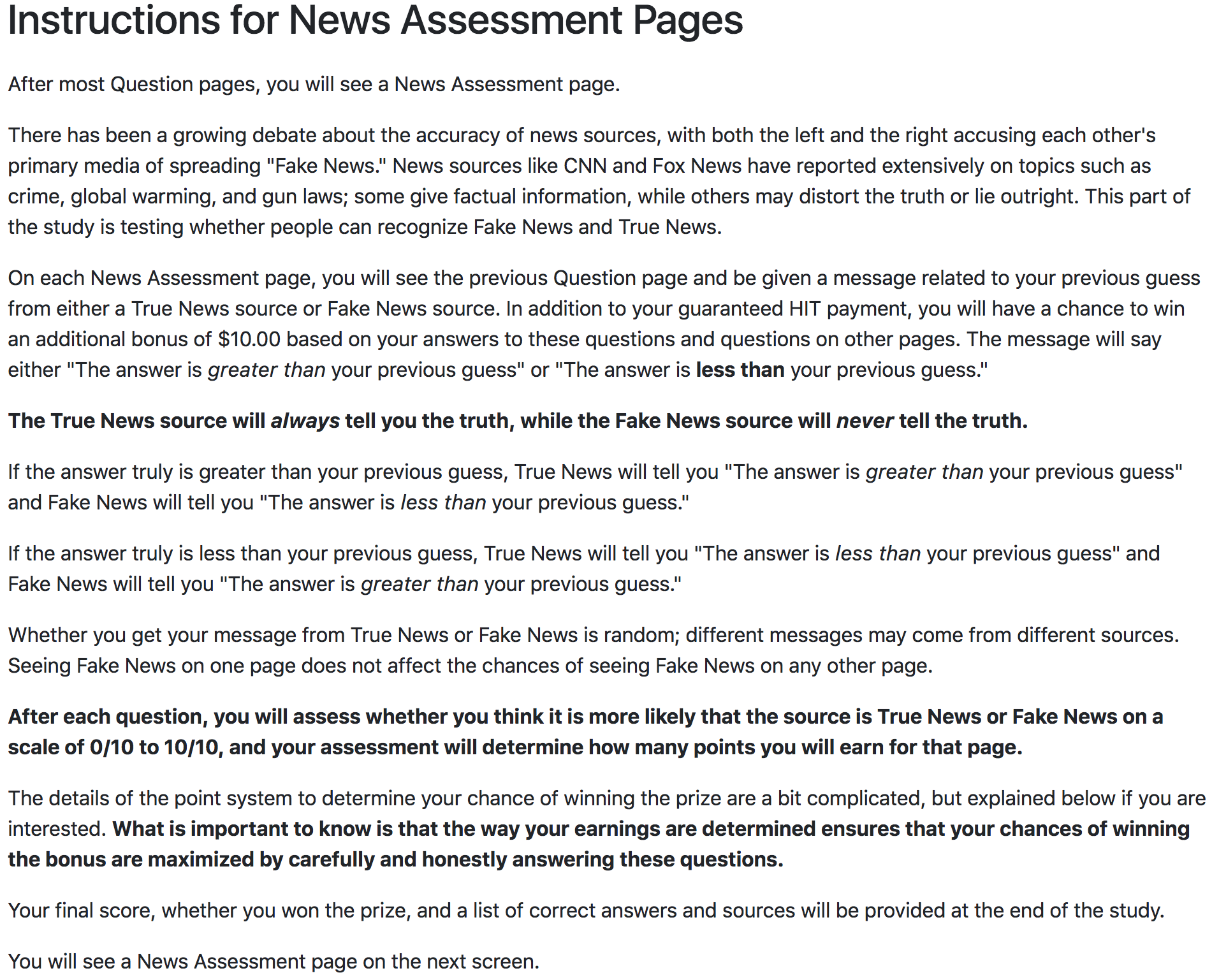}
\end{center}

\label{instructions_news_points}
\begin{center}
\includegraphics[width = \textwidth]{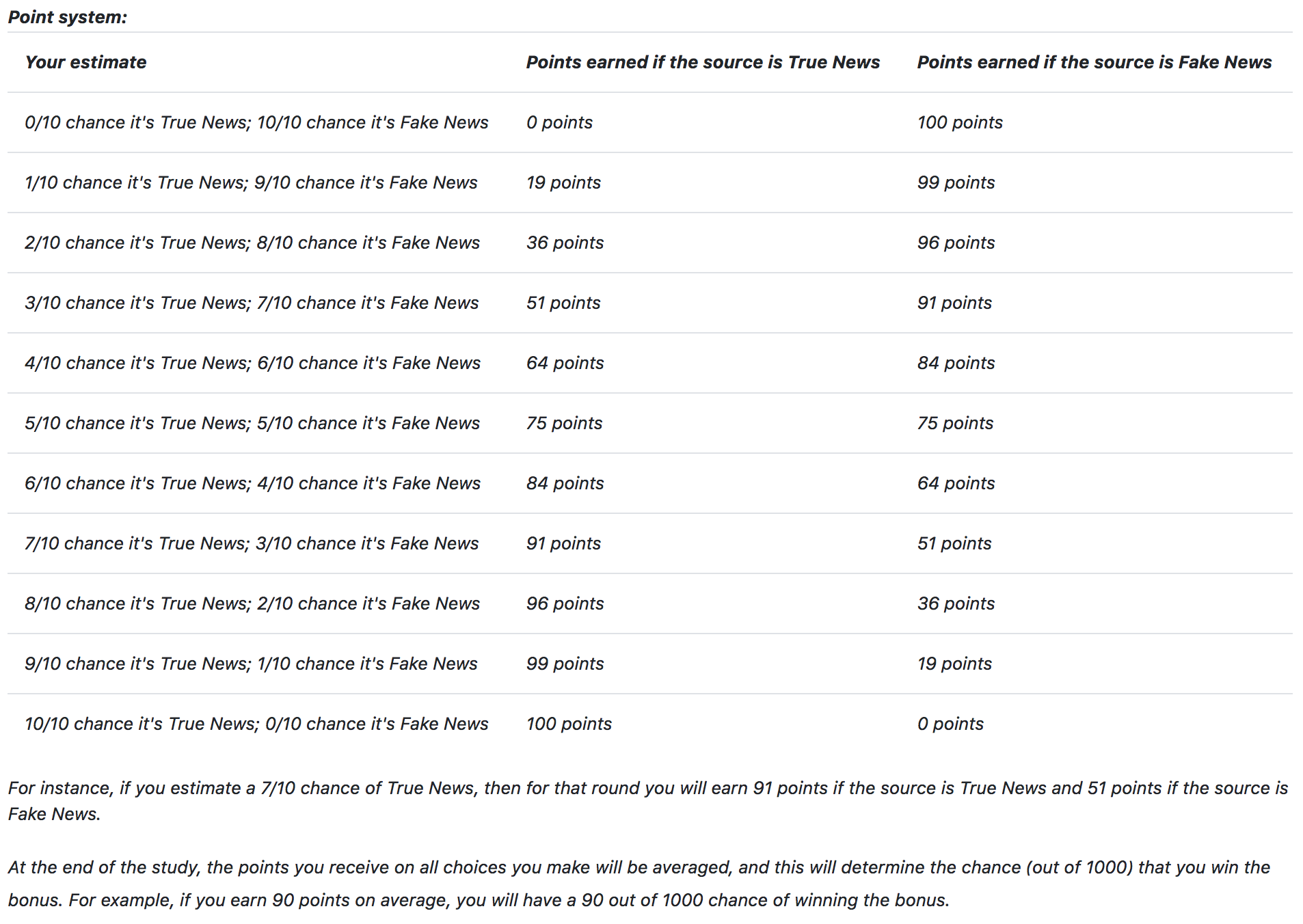}
\end{center}

\newpage

\begin{figure}
\begin{center}
\includegraphics[width = \textwidth]{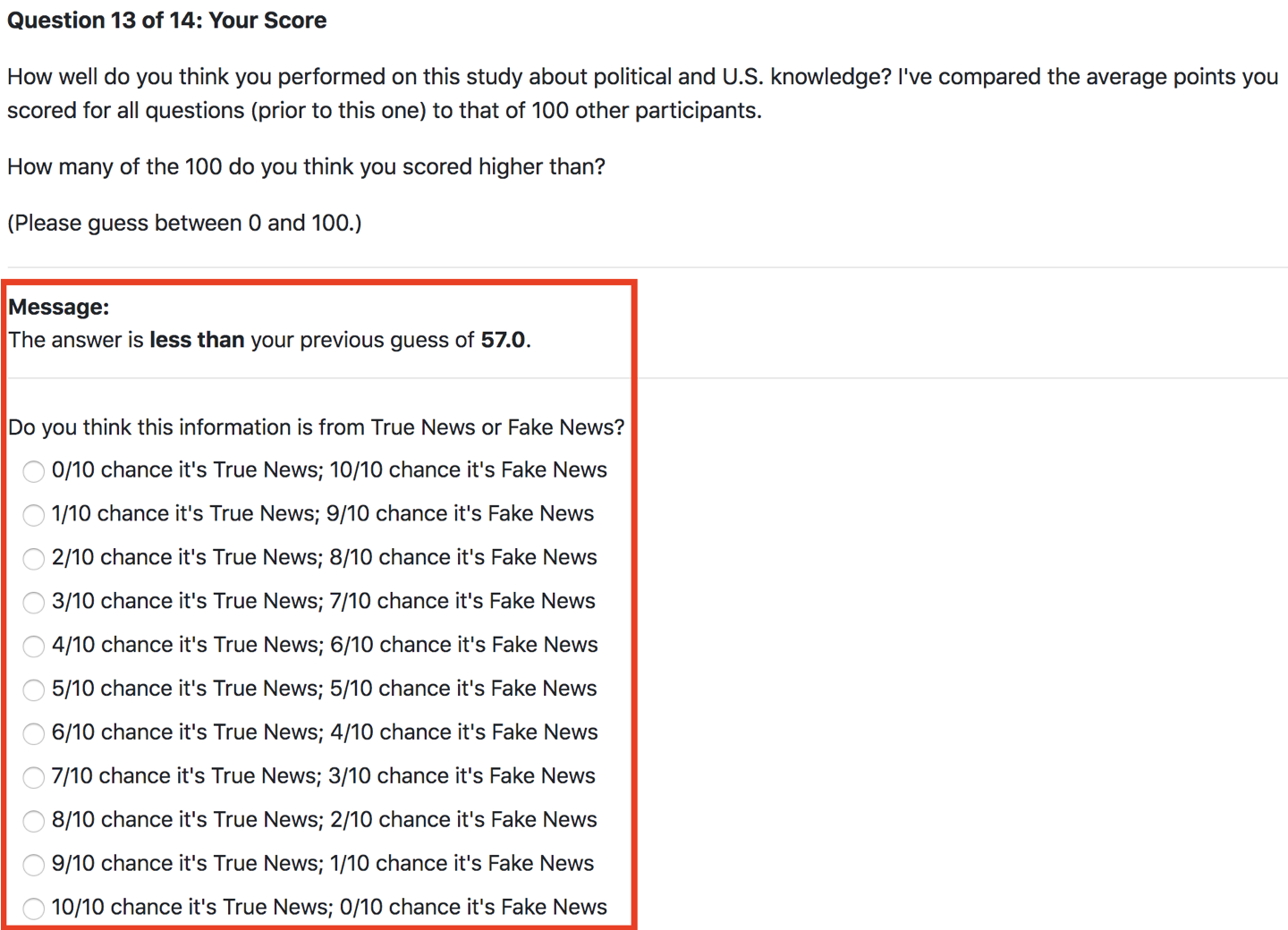}
\end{center}
\caption{The news assessment page for the performance question. The red box is to illustrate the message, and was not shown to participants.}
\label{news3}
\end{figure}

\clearpage

\begin{center}
\includegraphics[width = \textwidth]{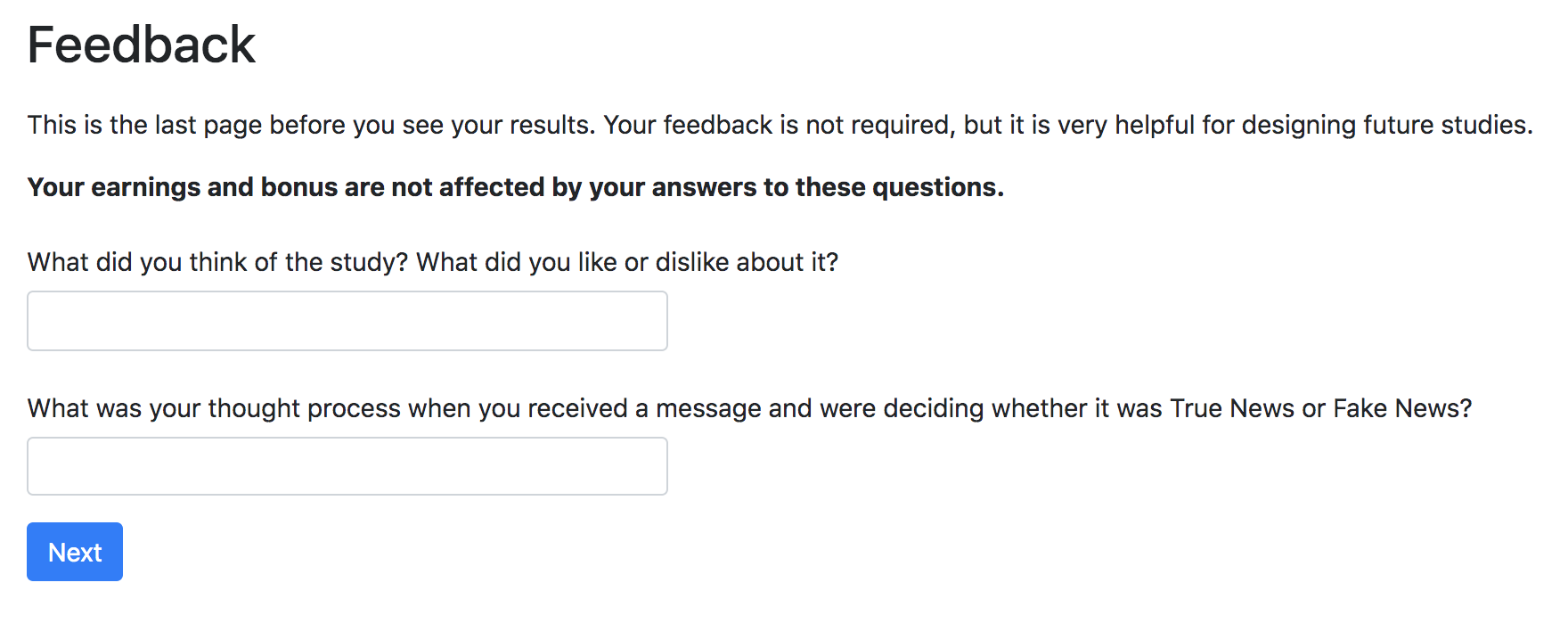}
\end{center}

\newpage

\begin{center}
\includegraphics[width = \textwidth]{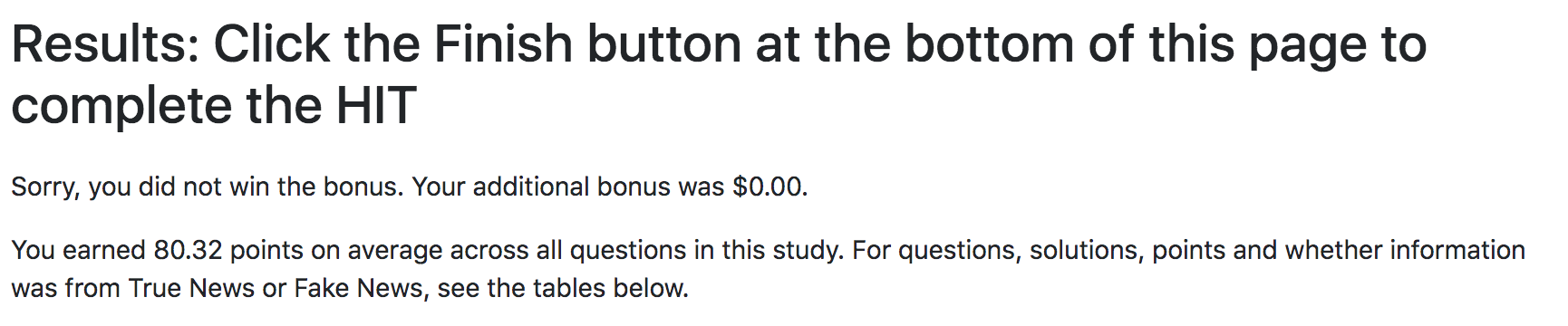}
\end{center}

\begin{center}
\includegraphics[width = \textwidth]{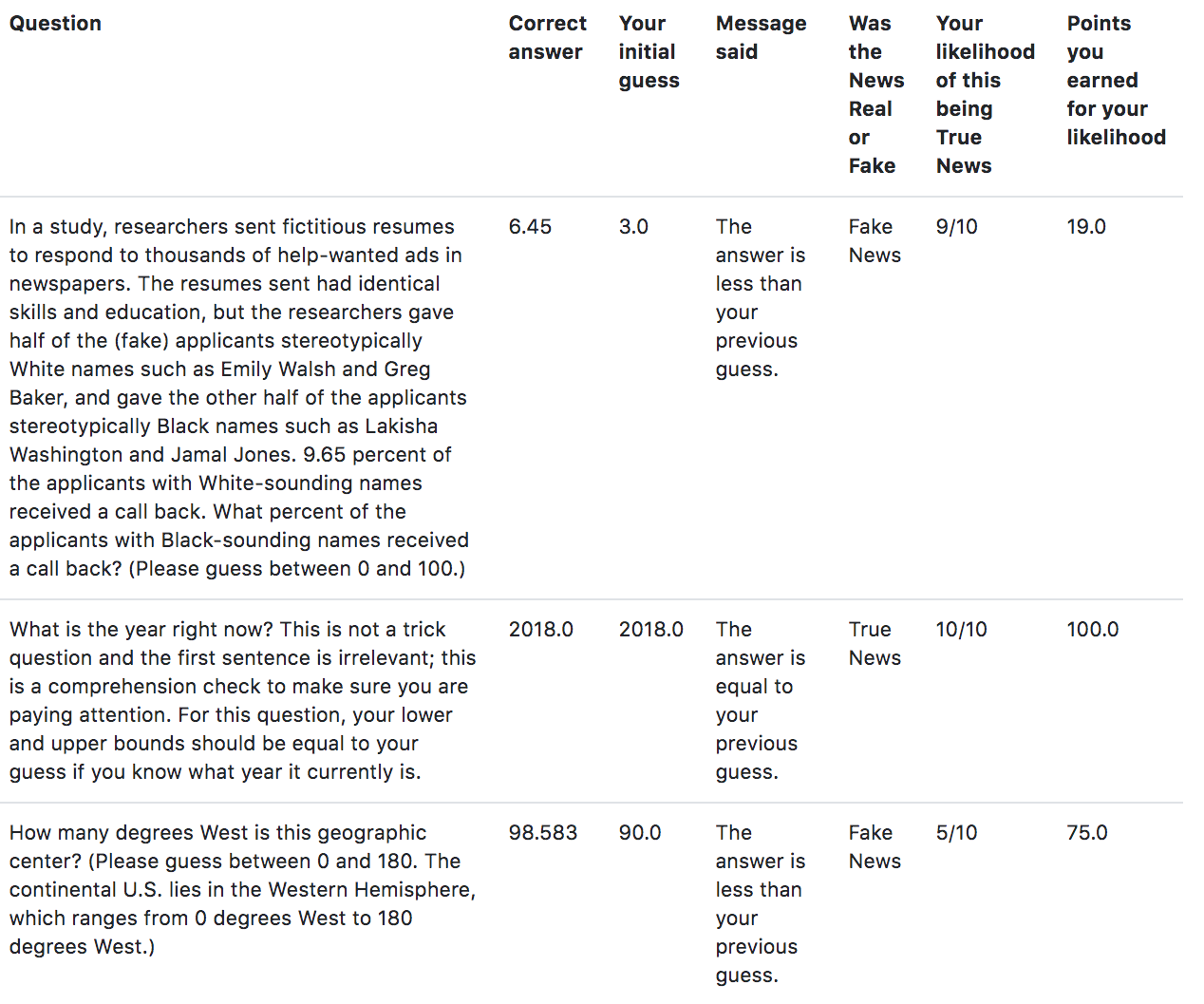}
\end{center}

\newpage

\begin{center}
\includegraphics[width = \textwidth]{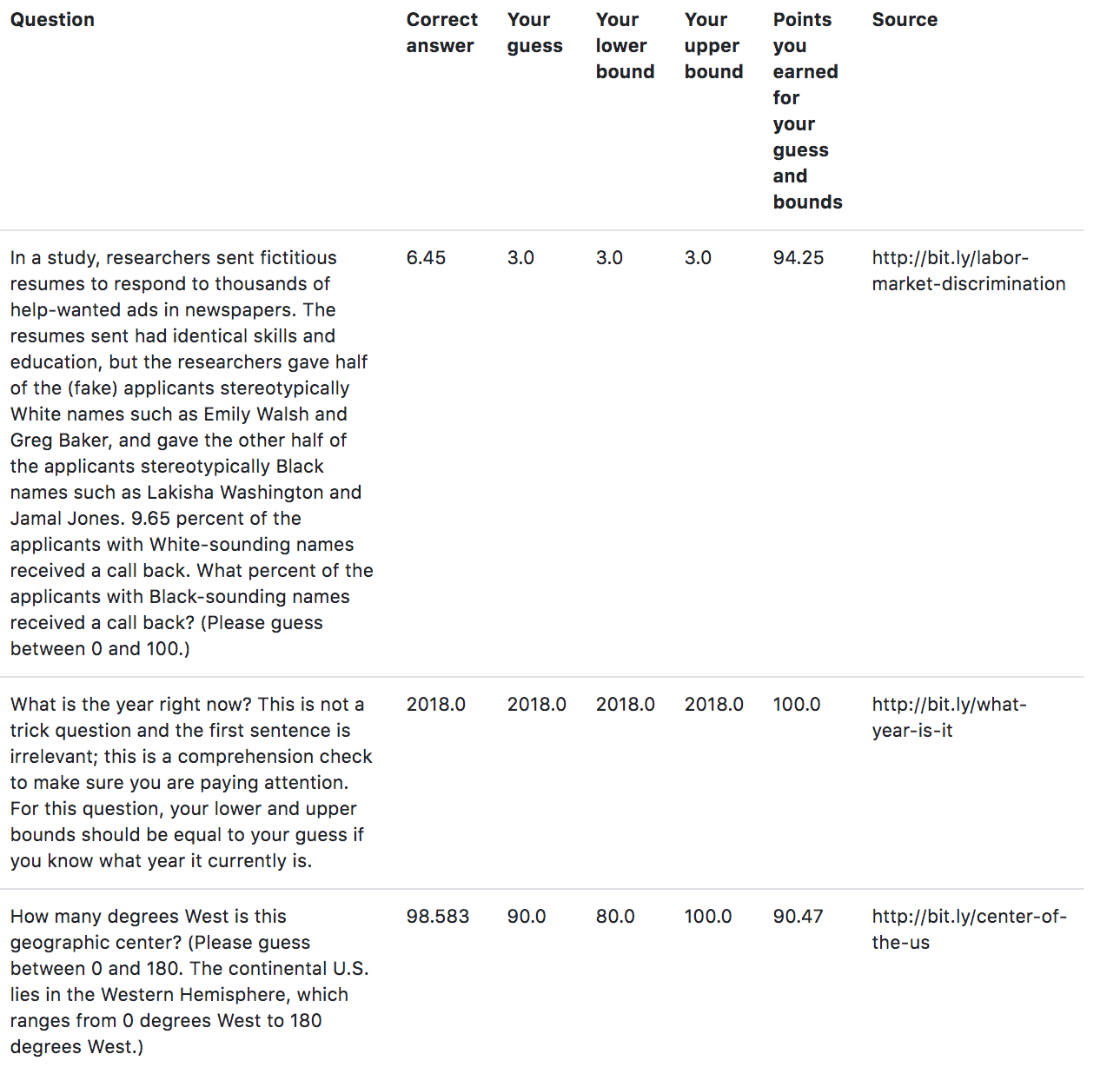}
\end{center}

\end{document}